\documentclass[conference]{IEEEtran}
\IEEEoverridecommandlockouts
\usepackage{cite}
\usepackage{amsmath,amssymb,amsfonts}
\usepackage{algorithmic}
\usepackage{graphicx}
\usepackage{textcomp}
\usepackage{balance}
\usepackage{breqn}
\usepackage[linesnumbered,ruled,vlined]{algorithm2e}
\usepackage{amsmath}
\usepackage{makecell}
\usepackage[table]{xcolor}
\newtheorem{definition}{Definition}
\usepackage{multicol}
\usepackage{multirow}
\usepackage{colortbl}
\newtheorem{theorem}{Theorem}
\newtheorem{corollary}{Corollary}

\usepackage[capitalize,noabbrev]{cleveref}
\usepackage{bm}
\usepackage[stable]{footmisc} 

\newcommand{\eat}[1]{}

\allowdisplaybreaks[4]

\def\BibTeX{{\rm B\kern-.05em{\sc i\kern-.025em b}\kern-.08em
    T\kern-.1667em\lower.7ex\hbox{E}\kern-.125emX}}
\begin{document}

\title{Privacy-Preserving Approximate Nearest Neighbor Search on High-Dimensional Data}

\author{
{
Yingfan Liu$^{\dagger}$, Yandi Zhang$^{\dagger}$, Jiadong Xie$^{\ddagger}$, Hui Li$^{\dagger,\S,\star}$, Jeffrey Xu Yu$^{\ddagger}$, Jiangtao Cui$^{\dagger}$}
\vspace{3.2mm}
\\
\fontsize{10}{10}
\selectfont\itshape
$^\dagger$School of Computer Science and Technology, Xidian University, Xi'an, China\\
$^\ddagger$The Chinese University of Hong Kong, Hong Kong SAR, China \\
$^\S$Shanghai Yunxi Technology, China / $^\star$Corresponding author
\fontsize{10}{10}
\selectfont\itshape
\\
\fontsize{9}{9} \selectfont\ttfamily\upshape
liuyingfan@xidian.edu.cn, ydzhang\_2@stu.xidian.edu.cn, jdxie@se.cuhk.edu.hk\\
hli@xidian.edu.cn, yu@se.cuhk.edu.hk, cuijt@xidian.edu.cn
}

\maketitle


\begin{abstract}
In the era of cloud computing and AI, data owners outsource ubiquitous vectors to the cloud, which furnish approximate $k$-nearest neighbors ($k$-ANNS) services to users. To protect data privacy against the untrusted server, privacy-preserving $k$-ANNS (PP-ANNS) on vectors has been a fundamental and urgent problem. However, existing PP-ANNS solutions fall short of meeting the requirements of data privacy, efficiency, accuracy, and minimal user involvement concurrently. To tackle this challenge, we introduce a novel solution that primarily executes PP-ANNS on a single cloud server to avoid the heavy communication overhead between the cloud and the user. To ensure data privacy, we introduce a novel encryption method named distance comparison encryption, facilitating secure, efficient, and exact distance comparisons. To optimize the trade-off between data privacy and search performance, we design a privacy-preserving index that combines the state-of-the-art $k$-ANNS method with an approximate distance computation method. Then, we devise a search method using a filter-and-refine strategy based on the index. Moreover, we provide the security analysis of our solution and conduct extensive experiments to demonstrate its superiority over existing solutions. Based on our experimental results, our method accelerates PP-ANNS by up to 3 orders of magnitude compared to state-of-the-art methods, while not compromising the accuracy.
\end{abstract}


\section{Introduction}
\label{sec:intro}

Approximate $k$-nearest neighbors ($k$-ANN) search ($k$-ANNS) on high-dimensional vectors has been a fundamental problem in various fields such as machine learning \cite{jordan2015machine}, information retrieval \cite{baeza1999modern} and retrieval-augmented generation \cite{RAG}. Let $P \subset \mathbb{R}^d$ be a database with $n$ vectors in $d$-dimensional space. Consider a $k$-ANNS query request denoted as ($\bm{q},k$), where $\bm{q} \in \mathbb{R}^d$ is a $d$-dimensional vector and $k$ is an integer parameter. The $k$-ANNS retrieves $k$ sufficiently close vectors in the database $P$ for the query $\bm{q}$.

In the era of cloud computing, data owners such as enterprises and organizations often opt to delegate the management of their vector databases and $k$-ANNS services to the cloud to cut down on data management costs and leverage scalable cloud resources. $k$-ANNS techniques applied to plaintext databases typically utilize index structures like locality-sensitive hashing~\cite{MPlsh}, inverted files~\cite{jegou2010product}, and proximity graphs~\cite{malkov2018efficient} to accelerate the search process.
However, the cloud is not entirely trusted and is typically considered \emph{honest-but-curious} \cite{zheng2024achieving, peng2017reusable}. As a result, conventional $k$-ANNS approaches are not directly applicable in cloud settings, as they risk exposing sensitive information, e.g., the database, query, and immediate search results, to the curious cloud. Consequently, there is an urgent need for privacy-preserving $k$-ANNS (PP-ANNS) solutions in such contexts.


\eat{
\begin{figure}[t]
	\centering
	\includegraphics[width=0.9\linewidth]{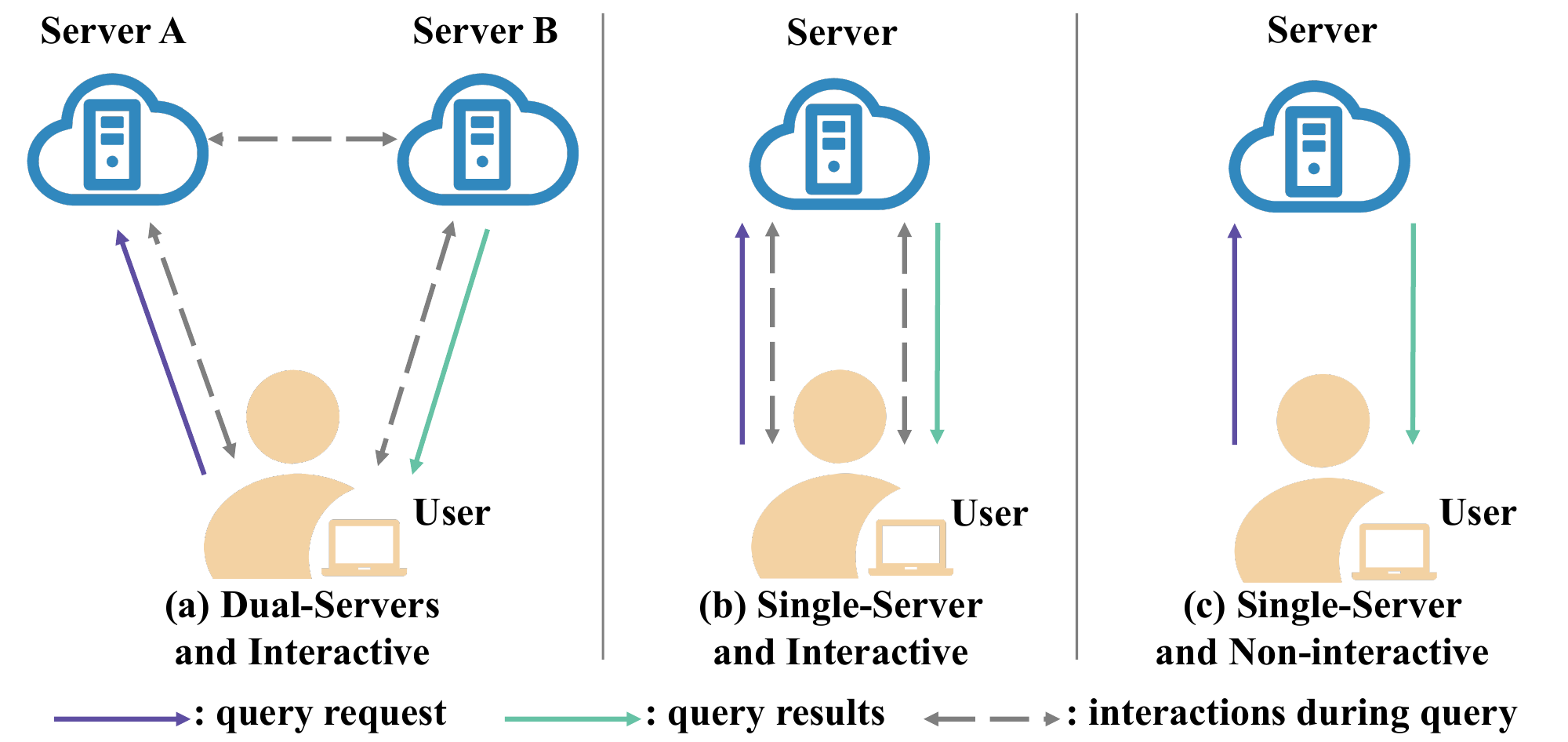}
	\caption{The query process of existing solutions}
	\label{fig:intro}
 \vspace{-3mm}
\end{figure}
}


As widely recognized~\cite{zheng2024achieving,servan2022private\eat{zhou2024pacmann},peng2017reusable}, a solution for PP-ANNS should effectively uphold the following three properties simultaneously:

\begin{enumerate}
    \item [$\bullet$]  \textbf{(P1) Privacy-preserving}: It must protect the privacy of database $P$ and the query $\bm{q}$ against the untrusted server.
    

    \item  [$\bullet$] \textbf{(P2) Efficient and Accurate}: 
    Efficiency and accuracy are the key aspects of $k$-ANNS performance. Hence, the solution should efficiently return high-quality results.


    \item  [$\bullet$] \textbf{(P3) Minimizing User Involvement and Interaction}: The user should not be involved in the search process except for encrypting queries and receiving results due to limited computing resources. 

    
\end{enumerate}

\vspace{1mm}
\noindent \textbf{Existing Solutions.}
To preserve data privacy, existing solutions~\cite{furon2013fast, servan2022private, zhou2024pacmann, wong2009secure, fuchsbauer2022approximate, zheng2024achieving} leverage encryption techniques to secure database vectors by storing them encrypted alongside an auxiliary index in the cloud. Those encryption techniques fall into two categories: distance incomparable encryption and distance comparable encryption. 
The former cannot directly compare the distances of vectors over their ciphertexts, while the latter can.
The encryption methods such as Advanced Encryption Standard~\cite{singh2013study} and Data Encryption Standard~\cite{singh2013study} belong to distance incomparable encryption methods, where the user retrieves a sufficient number of encrypted candidate vectors from the cloud using the index and subsequently computes their distances to the query after vector decryption. Nonetheless, such methods contend with substantial communication overhead between the cloud and the user, leading to inefficiencies and underutilization of cloud resources. On the other hand, distance comparable encryption methods, including asymmetric scalar-product-preserving encryption (ASPE) \cite{wong2009secure, zhu2013secure, zhang2018secure}, distance-comparison-preserving encryption (DCPE)~\cite{fuchsbauer2022approximate}, asymmetric matrix encryption (AME)~\cite{zheng2024achieving} and homomorphic encryption (HE)~\cite{zheng2018efficient, zheng2019achieving, guan2021toward} enable distance comparisons in the cloud, which further facilitates PP-ANNS service within the cloud upon receiving the encrypted query from the user. Unfortunately, ASPE and its variations fail to preserve data privacy as our analysis in Section~\ref{ssec:revisit_ASPE}, while DCPE produces inaccurate results. Both AME and HE suffer from huge computational costs. Therefore, none of the existing solutions satisfies all three properties concurrently.

\begin{figure}[t]
	\centering
	\includegraphics[width=0.85\linewidth]{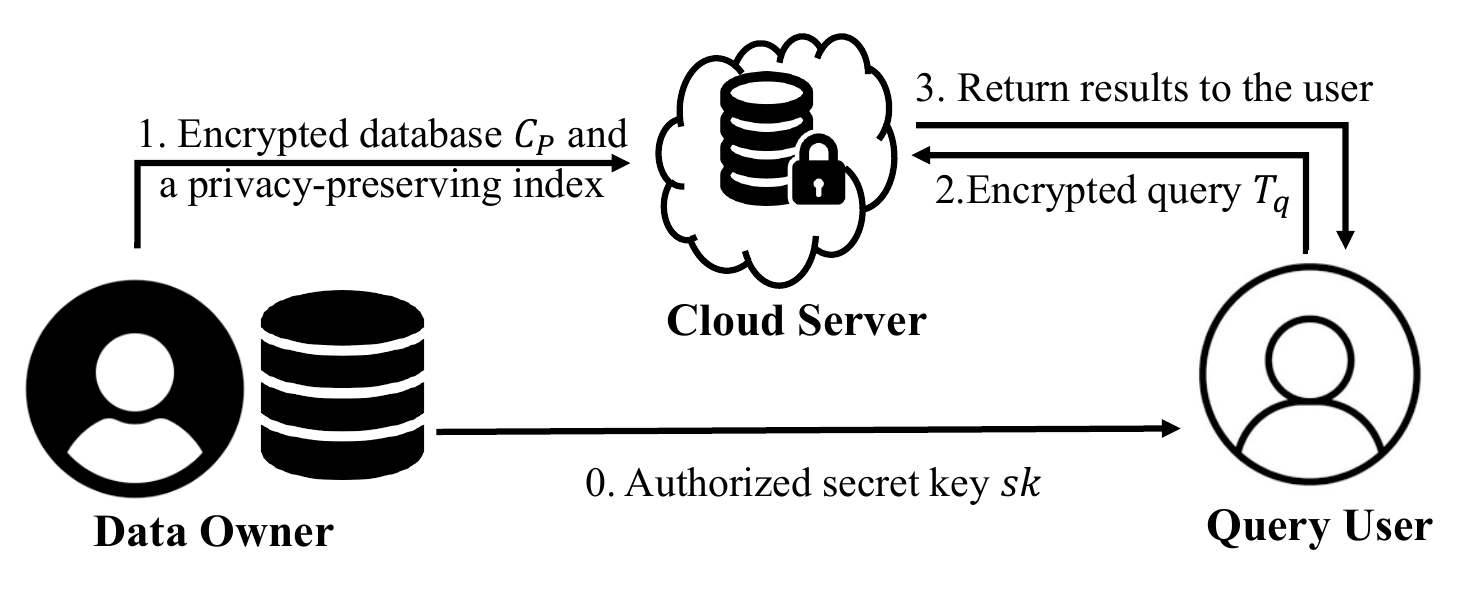}
 \vspace{-5mm}
	\caption{The system model of our PP-ANNS scheme.}
	\label{fig:sys_model}
 \vspace{-3mm}
\end{figure}

\vspace{1mm}
\noindent \textbf{Our Solution.}
%
To fulfill \textbf{P3}, we introduce a new PP-ANNS scheme with the system model shown in Figure~\ref{fig:sys_model}. The data owner outsources the encrypted database and a privacy-preserving index to the cloud. Throughout the search process, the user solely computes the encrypted query $T_{\bm{q}}$ and forwards it to the cloud, which in turn processes the query and transmits the results back to the user. 

To satisfy \textbf{P1}, we introduce a novel encryption technique named distance comparison encryption (DCE), which securely, exactly and efficiently answers distance comparisons over encrypted vectors. It comprises two phases: vector randomization and vector transformation. The first phase generates a random vector for the input vector through processes like random permutation, vector splitting, and the addition of random values, while the second phase converts the generated random vector into a DCE encrypted vector via operations such as element-wise vector manipulations and matrix encryption. We then provide theoretical proof of the correctness of distance comparison using the DCE method and analyze its efficiency and security.
To meet \textbf{P2}, integrating a privacy-preserving index into our DCE method is necessary to avoid scanning the entire database while returning accurate results. State-of-the-art $k$-ANNS techniques, proximity graphs such as hierarchical navigable small world (HNSW) \cite{malkov2018efficient}, notably outperform alternatives like locality-sensitive hashing and inverted files~\cite{survey2021}. 
A simple method is to deploy a proximity graph such as HNSW directly in the cloud and conduct $k$-ANNS on the graph with distance comparisons on DCE encrypted vectors. 
However, such a naive method causes two issues, i.e., (1) the risk of exposing sensitive information since the edges in the graph indicate the relationships between vectors, and (2) low efficiency since each DCE computation costs at least $4\times$ that of a normal distance computation as analyzed in Section~\ref{ssec:dce}.
To enhance data privacy, we design a privacy-preserving index that constructs an HNSW graph over the encrypted database using distance-comparison-preserving encryption (DCPE)~\cite{fuchsbauer2022approximate}, instead of the plaintext database. While DCPE is secure and efficient~\cite{fuchsbauer2022approximate}, it answers only approximate distances, and thus conducting $k$-ANNS over the DCPE encrypted vectors will decrease the accuracy even using HNSW as the index. 
To balance efficiency and accuracy, we develop a search approach based on the \emph{filter-and-refine} strategy. At a high level, in the \emph{filter} phase, the $k$-ANN search with HNSW on DCPE encrypted database yields a selected group of top-quality candidates, where distances are computed over the DCPE encrypted vectors. Note that such a distance computation costs exactly the same as the normal distance computation. Next, in the \emph{refine} phase, the server conducts exact distance comparisons based on our DCE method among these candidates to identify the best $k$ vectors. In this way, we reduce the number of costly distance comparisons of DCE while not compromising the accuracy. 

\vspace{1mm}
\noindent \textbf{Contributions.}
Our main contributions are as follows.

\begin{itemize}
    \item We introduce a novel encryption method, DCE, facilitating secure, exact, and efficient distance comparisons among vectors.
    \item We design a privacy-preserving index and introduce a \emph{filter-and-refine} search strategy based on this index.
    \item Equipped with the above techniques, we present a novel PP-ANNS scheme that upholds all three properties.
    \item We conduct a theoretical security analysis of our proposed PP-ANNS scheme and perform comprehensive experiments on real-world datasets to showcase the superiority of our approach over state-of-the-art methods. Our experimental results demonstrate that our approach can be up to $1000\times$ faster than state-of-the-art methods while maintaining comparable accuracy.
\end{itemize}

\vspace{1mm}
\noindent \textbf{Organization.}
The remainder of this paper is organized as follows. We present the preliminaries in~\cref{sec:preli} and revisit the distance comparable encryption methods in~\ref{sec:revisit}. In~\cref{sec:secure_dc}, we show our distance comparable encryption method DCE. \cref{sec:ours} describes our PP-ANNS scheme. We present the security analysis in~\cref{sec:secana} and performance evaluation in~\cref{sec:exp}, respectively. \cref{sec:rel} introduces the related works. Finally, we conclude this paper in~\cref{sec:conclu}.

\section{Preliminaries}
\label{sec:preli}

In this section, we first present our system model and threat model, respectively. Afterward, we formally define the problem and present the challenges of our problem.

\eat{
\begin{table}[t]
  \caption{The definition of key notations}
  \label{tab:notation_table}
  \begin{tabular}{lc}
    \toprule
    Notations & Definition \\
    \midrule
    $P,\ P_i$ & The database and a vector in it \\
    $q$       & The query vector   \\
    $dist(\cdot,\cdot)$ & The Euclidean distance of two vectors \\
    $\circ$   & The vector multiplication \\
    $\times$  & The matrix multiplication \\
    $\cdot$   & The number multiplication \\
    $/$       & The vector division  \\
    
    \bottomrule
  \end{tabular}
\end{table}
}

\subsection{System Model}\label{ssec:sysmod}

In this work, we focus on the single-server and non-interactive scenario, where there are three types of participants in the system, i.e., data owner, user, and cloud server.

\sloppy
\resizebox{6pt}{!}{\raisebox{-1pt}\textbullet} \textbf{Data Owner}: Data owner owns a database $P \subset \mathbb{R}^d$, which includes $n$ $d$-dimensional vectors. Due to the limited capability of computation and storage, data owner encrypts each vector $\bm{p} \in P$ into ciphertexts $C_{\bm{p}}$, which is then outsourced to the cloud server. 
For simplicity, we define $C_P = \{C_{\bm{p}} | \bm{p} \in P\}$.

\resizebox{6pt}{!}{\raisebox{-1pt}\textbullet} \textbf{User}: User enjoys the query service by giving a query request with $(\bm{q},k)$, where $\bm{q} \in \mathbb{R}^d$ is a query vector and $k \in \mathbb{N}^+$ is the number of returned neighbors. To preserve the privacy of $\bm{q}$, the user encrypts $\bm{q}$ into ciphertext $T_{\bm{q}}$ that is then sent to the cloud server to answer the query.

\resizebox{6pt}{!}{\raisebox{-1pt}\textbullet} \textbf{Cloud Server}: Cloud server stores the encrypted database $C_P$ and responds to query requrests from user. When receiving $(T_{\bm{q}},k)$, it searches on $C_P$ to retrieve $k$ close neighbors and then returns to the user. 


\subsection{Threat Model}

In this work, the data owner is trusted, and the user is assumed to be honest. Both of them will honestly use the $k$-ANNS service, while not leaking their keys to attackers or colluding with the server. On the other hand, the server is considered to be \emph{honest-but-curious}. It means that the server will sincerely store the encrypted database for the data owner and offer the $k$-ANNS service to the user strictly following the predefined scheme. However, it is curious about sensitive information such as the database and queries. In this work, we aim to prevent the server from obtaining the database, the query vector, and immediate distance values during the search process. 
Besides, in line with the majority of secure query schemes with server-side index~\cite{peng2017reusable, servan2022private}, we compromise the information revealed by the index for the sake of the efficiency and accuracy of PP-ANNS. Although the index is built on pure ciphertexts, there still exists some information leaked by the index, e.g., the neighborhood relationship of vectors. Except that, nothing more should be revealed. 
As to other active attacks such as side-channel attacks, DOS attacks, and data pollution attacks, we leave them in our future works. 


\subsection{Problem Definition and Challenges}\label{ssec:prodef}

Based on our system model and threat model, we formally define the privacy-preserving $k$-ANN search as follows. 

\begin{definition}[\textbf{PP-ANNS}]
\label{def:PP-ANNS} 
Under our system model, given an encrypted query $T_{\bm{q}}$ and an integer $k$, privacy-preserving $k$-ANN search (PP-ANNS) problem aims to return approximate $k$ neighbors that are sufficiently close to the query vector $\bm{q}$ by searching the encrypted database $C_P$, without leaking $P$ or $\bm{q}$ to cloud server under our threat model.
\end{definition}

In this work, we use Euclidean distance to measure the distance between two vectors. Given a vector $\bm{p} \in P$ and a query vector $\bm{q}$, the squared Euclidean distance between them is denoted as {$dist(\bm{p}, \bm{q}) = ||\bm{p}, \bm{q}||^2 = \sum_{i=1}^d (p_i - q_i)^2 $}. 

Based on the threat model, we advocate that a secure solution to PP-ANNS should satisfy the \eat{selectively secure } \underline{{ind}}istinguishability under \underline{{k}}nown-\underline{{p}}laintext \underline{{a}}ttack (IND-KPA)~\cite{cambareri2015known} in the real/ideal world security model. Let $\mathcal{L}$ be the leaked information, the ideal world involves a probabilistic polynomial time adversary $\mathcal{A}$ and a simulator with leakage $\mathcal{L}$. The real world involves a probabilistic polynomial time adversary $\mathcal{A}$ and a challenger. We can formally define the IND-KPA security as follows.
\begin{definition}[\textbf{IND-KPA Security}]
\label{def:ind-kpa_security} 
A scheme is \eat{selectively secure}IND-KPA secure with leakage $\mathcal{L}$ iff for any $\mathcal{A}$ issuing a polynomial number of interactions, there exists a simulator such that the advantage that $\mathcal{A}$ can distinguish the views of real and ideal experiments is negligible, i.e., $\left|\mathrm{Pr}[\mathrm{View}_{\mathcal{A},\mathrm{Real}}=1]-\mathrm{Pr}\left[\mathrm{View}_{\mathcal{A},\mathrm{Ideal}}=1\right]\right|$ is negligible.
\end{definition}

{
To design a solution to PP-ANNS, as we previously discussed in the introduction, we must address two primary challenges: (1) how to securely, exactly, and efficiently conduct distance comparisons, and (2) how to reduce the number of distance comparisons during $k$-ANNS. Notably, secure, exact, and efficient distance comparisons are fundamental operations in PP-ANNS. In tackling the first challenge, we first revisit existing methods and ascertain that none achieve the trifecta of security, exactness, and efficiency concurrently in Section~\ref{sec:revisit}. Motivated by this, we introduce our DCE scheme in Section~\ref{sec:secure_dc}. In addressing the second challenge, we propose a privacy-preserving index that integrates the state-of-the-art $k$-ANNS method HNSW, and an approximate distance comparison encryption method DCPE, which strikes a balance between data privacy and search performance, in Section~\ref{sec:ours}.
}

\section{Revisiting Secure Distance Comparison}
\label{sec:revisit}

In this section, we revisit existing secure distance comparison (SDC) methods to determine that none of them achieves security, precision, and efficiency simultaneously.

SDC on the server is an essential operation in PP-ANNS. Existing SDC methods fall into two categories: those that compare distances through distance computation and those that perform direct distance comparisons. The former calculates distances over encrypted vectors to compare them, which contains methods based on asymmetric scalar-product-preserving encryption (ASPE)~\cite{furon2013fast, li2019insecurity, miao2023efficient, yuan2017practical} and distance-comparison-preserving encryption (DCPE)~\cite{fuchsbauer2022approximate}.
While the latter directly yields the result of the distance comparison without intermediate computations, which includes asymmetric matrix encryption (AME)~\cite{zheng2024achieving} and homomorphic encryption (HE)~\cite{xue2017secure, furon2013fast}.
Note that, in the analysis in this section, we exclude HE-based methods due to their significant computational overhead~\cite{zheng2024achieving}.

\subsection{ASPE Schemes Revisited} 
\label{ssec:revisit_ASPE}



First, we briefly explain the ASPE scheme~\cite{wong2009secure}. 
Given two vectors $\bm{p}, \bm{q} \in \mathbb{R}^d$ and an invertible matrix $M \in \mathbb{R}^{d \times d}$, we get the ciphertexts of vectors, i.e., $\texttt{Enc}(\bm{p})=(\bm{p}^T \times M)^T$ and $\texttt{Enc}(\bm{q})=M^{-1} \times \bm{q}$. The inner product $\bm{p}^T\bm{q}$ can be recovered by $\texttt{Enc}(\bm{p})^T \times \texttt{Enc}(\bm{q})$, i.e., the inner product of their corresponding encrypted vectors. 
Even though ASPE and its variants were created for the inner product, we can convert the squared Euclidean distance $||p, q||^2$ to the inner product of two new vectors, i.e., $\bm{p}^{\prime} = [\bm{p}, 1, ||\bm{p}||^2]^T$ and $\bm{q}^{\prime} = [-2\bm{q}, ||\bm{q}||^2, 1]^T$. 

Unfortunately, ASPE has been demonstrated to be KPA insecure for Euclidean distance and inner product~\cite{yao2013secure,lin2017revisiting,li2019insecurity}.
Hence, several ASPE variants were proposed to strengthen its security~ \cite{cao2013privacy, wang2014privacy} via specific transformations of distance values. 
In the following, we revisit the security of these ASPE variants and prove that they are not KPA secure. 




To be formal, we present the assumption under KPA. Let $P$ be the database and $Q$ be a set of queries. Given the encrypted database $C_P=\{C_{\bm{p}} | \bm{p} \in P\}$ and the encrypted query set $C_Q=\{ T_{\bm{q}} | \bm{q} \in Q \}$.
We denote  $L(C_{\bm{p}}, T_{\bm{q}})$ as the information about $dist(\bm{p}, \bm{q})$ leaked from the ciphertexts, where $\bm{p} \in P, \bm{q} \in Q$. 
We assume that the attacker owns $C_Q$, $C_P$, and $P_{leak} \subset P$, where $|P_{leak}| = m$. 
$L(C_{\bm{p}}, T_{\bm{q}})$ in ASPE and its enhanced schemes is a specific transformation of $dist(\bm{p}, \bm{q})$, such as linear~\cite{cullen2012matrices}, exponential~\cite{barndorff1982exponential}, logarithmic~\cite{keene1995log}, and square~\cite{bailey2004efficient}. 
Then, we prove that those additional transformations cannot prevent the attacker from recovering the plaintexts of $P$ and $Q$.


\begin{theorem}
\label{lemma:linear_trans}
The enhanced ASPE scheme that leaks \textbf{the linear transformation of distances} is not KPA secure. 
\end{theorem}

\begin{IEEEproof}
\sloppy
Given $C_Q$, $C_P$ and $P_{leak}$, to recover a query vector $\bm{q}$, attacker constructs $d+2$ equations according to ASPE with linear transformation, $[-2\bm{p}_i^T, ||\bm{p}_i||^2, 1] \times [r_1\bm{q}^T, r_1, r_2]^T = L\small(C_{\bm{p}_i}, T_{\bm{q}}\small)$, where $\bm{p}_i \in P_{leak}$, $1 \leq i \leq d+2$ and $r_1, r_2$ are random numbers. 
Those $d+2$ equations could be reformed as $M_c \bm{x} = \bm{b}$, where the $i$-th ($1\leq i \leq d+2$) row of matrix $M_c \in \mathbb{R}^{(d+2)\times (d+2)}$ is $\bm{p}_i^T$, the vector $\bm{b} = \big [L(C_{\bm{p}_i}, T_{\bm{q}}), \cdots, L(C_{\bm{p}_{d}}, T_{\bm{q}_{d+2}}) \big]^T$ and $\bm{x} =[r_1\bm{q}^T, r_1, r_2]^T$. 
When $M_c$ is invertible, $\bm{x} = M_c^{-1}\bm{b}$ and thus $\bm{q}$ is recovered. Similarly, with $d+2$ recovered queries, to recover any database vector $\bm{p} \in P \setminus P_{leak}$, attacker constructs another $d+2$ equations in a similar way, $[-2\bm{p}^T, ||\bm{p}||^2, 1] \times [r_{1j}\bm{q}_j^T, r_{1j}, r_{2j}]^T  = L\small(C_{\bm{p}}, T_{\bm{q}_j}\small)\ (1 \leq j \leq d+2)$. Similarly, $\bm{p}$ could be recovered. To sum up, the enhanced ASPE scheme that leaks the linear transformation of distances is not secure. 
\end{IEEEproof}

\begin{corollary}
\label{lemma:exp_trans}
The enhanced ASPE scheme that leaks \textbf{the exponential transformation of distances} is not KPA secure. 
\end{corollary}
\begin{IEEEproof}
\sloppy
Given $C_Q$, $C_P$ and $P_{leak}$, to recover a query vector $\bm{q}$, attacker constructs $d+2$ equations according to ASPE with exponential transformation, $[-2\bm{p}_i^T, \|\bm{p}_i\|^2, 1] \times [r_{1}\bm{q}^T, r_{1}, r_{2}]^T = \ln L\small(C_{\bm{p}_i}, T_{\bm{q}}\small)$, where $\bm{p}_i \in P_{leak}$, $1 \leq i \leq d+2$ and $r_{1},r_{2} \in \mathbb{R}$ are two random numbers. 
It is converted to \cref{lemma:linear_trans} by replacing $L(C_{\bm{p}_i}, T_{\bm{q}})$ with its logarithmic value. Following the same idea of \cref{lemma:linear_trans}, each query vector could be recovered. Then, with $d+2$ recovered queries, each database vector can be recovered in the same way. 
\end{IEEEproof}

\begin{corollary}
\label{lemma:log_trans}
The enhanced ASPE scheme that leaks \textbf{the logarithmic transformation of distances} is not KPA secure. 
\end{corollary}

\begin{IEEEproof}
 \sloppy
Given $C_Q$, $C_P$ and $P_{leak}$, to recover a query vector $\bm{q}$, attacker constructs $d+2$ equations according to ASPE with logarithmic transformation, $[-2\bm{p}_i^T, \|\bm{p}_i\|^2, 1] \times [r_{1}\bm{q}^T, r_{1}, r_{2}]^T = e^{L\small(C_{\bm{p}_i}, T_{\bm{q}}\small)}$, where $\bm{p}_i \in P_{leak}$, $1 \leq i \leq d+2$ and $r_{1},r_{2} \in \mathbb{R}$ are two random numbers. 
It is converted to \cref{lemma:linear_trans} by replacing $L(C_{\bm{p}_i}, T_{\bm{q}})$ with its exponential value. Following the same idea of \cref{lemma:linear_trans}, the query vector $\bm{q}$ could be recovered. In the same way, with $d+2$ recovered queries, each database vector can be recovered. 
\end{IEEEproof}

\begin{theorem}
\label{lemma:squad_trans}
The enhanced ASPE scheme that leaks \textbf{the square of distances} is not KPA secure. 
\end{theorem}
\begin{IEEEproof}
\sloppy
Given $C_Q$, $C_P$ and $P_{leak}$, to recover a query vector $\bm{q}$, according to the ASPE scheme with the square of distances, we have $L(C_{\bm{p}_i}, T_{\bm{q}}) = ([-2\bm{p}_i^T, \|\bm{p}_i\|^2, 1] \times [r_{1}\bm{q}^T, r_{1}, r_{2}]^T)^2$. By extending this equation, we have
\begin{equation*}
\begin{aligned}
L(C_{\bm{p}_i}, T_{\bm{q}}) =&r_{1}\cdot (\|\bm{p}_i\|^{2}-2\cdot \bm{p}_i^T\bm{q}+r_{2})^{2}+r_{3}\\
	=&r_{1}\cdot (\|\bm{p}_i\|^{4}-4\|\bm{p}_i\|^{2}\cdot \bm{p}_i^T\bm{q} + 2\|\bm{p}_i\|^{2}\cdot r_{2}\\ 
    &+4\cdot (\bm{p}_i^T\bm{q})^2
     - 4\cdot \bm{p}_i^T\bm{q} \cdot r_{2}+r_{2}^{2})+r_3.\\
\end{aligned}    
\end{equation*}
For each vector $\bm{p} = [p_1, p_2, \cdots, p_d]^T$, $\bm{p}^2 = [p_1^2, p_2^2, \cdots, p_d^2]^T$ and $\ddot{\bm{p}}^{-i} = [p_1, \cdots, p_{i-1}, p_{i+1}, \cdots, p_d]^T$. Here, $\ddot{\bm{p}}^{-i} \in \mathbb{R}^{d-1}$ is obtained by just removing the $i$-th coordinate of $\bm{p}$. Then, $L(C_{\bm{p}_i}, T_{\bm{p}})$ could be represented as the inner product of two new vectors $\bm{p}_i^{\prime}, \bm{q}^{\prime} \in \mathbb{R}^{0.5d^2+2.5d+3}$, where 
$$\begin{aligned}
	&\bm{p}^{\prime}_i=[\|\bm{p}_i\|^{4},\ \|\bm{p}_i\|^{2}\cdot \bm{p}_i, \|\bm{p}_i\|^{2}, 4 \cdot \bm{p}^2, 8p_{i1} \cdot \ddot{\bm{p}_i}^{-1}, 8p_{i2} \cdot \ddot{\bm{p}_i}^{-2}, \\
          &\quad\quad \cdots, 8p_{id} \cdot \ddot{\bm{p}_i}^{-d},-4 \cdot \bm{p}_i, 1]^T,\\
	&\bm{q}^{\prime}=[r_{1},\  -4r_1 \cdot \bm{q},\ 2r_2,  r_{1} \cdot \bm{q}^2, r_{1}q_{1} \cdot \ddot{\bm{q}}^{-1}, r_{1}q_{2} \cdot \ddot{\bm{q}}^{-2}, \\
    &\quad\quad \cdots, r_{1}q_{d} \cdot \ddot{\bm{q}}^{-d}, r_{1}r_{2}\cdot \bm{q}, r_{1}r_{2}^{2}+r_{3}]^T.
\end{aligned}$$
Here, $\bm{p}_i = [p_{i1}, \cdots, p_{id}]^T$.Given $T_{\bm{q}}$, $C_P$ and $P_{leak}$, to recover a query vector $\bm{q}$, attacker constructs $0.5d^2+2.5d+3$ equations via the above two new vectors, ${\bm{p}^{\prime}_i}^T{\bm{q}^{\prime}} = L(C_{\bm{p}_i}, T_{\bm{q}})$, where $1 \leq i \leq 0.5d^2+2.5d+3$. 
Following the same idea of \cref{lemma:linear_trans}, $\bm{q}^{\prime}$ could be recovered, and thus $\bm{q}$ could be recovered. 
Again, with $0.5d^2+2.5d+3$ recovered queries, to recover each database vector $\bm{p} \in P \setminus P_{leak}$, the attacker constructs $0.5d^2+2.5d+3$ equations as the same procedure.
Hence, the scheme that leaks the distances is not secure. 
\end{IEEEproof}

Moreover, other existing APSE variants~\cite{miao2023efficient,yuan2017practical,kim2018function} could be aligned with the aforementioned cases in this part. 
\emph{In conclusion, ASPE schemes that reveal specific transformations of distances are not KPA secure.}

\subsection{DCPE Scheme Revisited}
\label{ssec:dcpe}

Unlike ASPE-based methods that reveal the distance values or their transformations, distance-comparison-preserving encryption~(DCPE) only reveals approximate distance values by the $\beta$-approximate-distance-comparison-preserving ($\beta$-DCP) function~\cite{fuchsbauer2022approximate}. 
\begin{definition}
    (\textbf{$\beta$-DCP}) For $\beta \in \mathbb{R}^+$, a $\beta$-DCP function $f: \mathbb{R}^d \to \mathbb{R}^d$ satisfies that $\forall{\bm{o}, \bm{p}, \bm{q} \in \mathbb{R}^d}$, if $dist(\bm{o}, \bm{q}) < dist(\bm{p}, \bm{q}) - \beta$, then $dist(f(\bm{o}), f(\bm{q})) < dist(f(\bm{p}), f(\bm{q}))$. 
\end{definition}
\eat{
\begin{definition}
    (\textbf{$\beta$-DCP}) For $\beta \in \mathbb{R}^+$, a function $f: \mathbb{R}^d \to \mathbb{R}^d$ is said to be $\beta$-DCP, if $\forall{\bm{o}, \bm{p}, \bm{q} \in \mathbb{R}^d}$, 
    \begin{equation*}
        \begin{split} 
        \forall{\bm{o}, \bm{p}, \bm{q} \in \mathbb{R}^d},\quad &dist(\bm{o}, \bm{q}) < dist(\bm{p}, \bm{q}) - \beta \\
        \Rightarrow \quad &dist(f(\bm{o}), f(\bm{q})) < dist(f(\bm{p}), f(\bm{q}))
        \end{split}
    \end{equation*}
\end{definition}
}

Let $f(P) = \{f(\bm{p}) | \bm{p} \in P\}$. Let $\bm{u}^*$ be the exact nearest neighbor of $f(\bm{q})$ on $f(P)$. The authors in \cite{fuchsbauer2022approximate} claim that $dist(\bm{q}, \bm{u}^*) \leq dist(\bm{q}, \bm{p}) + \beta$ holds for each $\bm{p} \in P$. Besides, they propose such a $\beta$-DCP function which first scales each vector $\bm{p} \in P$ with the same scaling factor and then adds a random vector to $\bm{p}$. As proven in \cite{fuchsbauer2022approximate}, DCPE is IND-KPA secure. Notabley, the encrypted vector still has $d$ dimensions, and $dist(f(\bm{p}), f(\bm{q}))$ costs the same as $dist(f(\bm{o}), f(\bm{q}))$. Hence, DCPE is efficient, but only returns approximate distance values. PP-ANNS with DCPE cannot ensure the accuracy. 

\subsection{AME Scheme Revisited}
\label{ssec:ame_revisited}
Different to ASPE and DCPE, which reveal information about distance values, asymmetric matrix encryption (AME)~\cite{zheng2024achieving} only reveals the results of distance comparisons. 
As in \cite{zheng2024achieving}, AME is KPA secure, but pretty costly. 
To be specific, AME generates the secret key consisting of 32 matrices in $\mathbb{R}^{(2d+6) \times (2d+6)}$. With those matrices, each database vector is encrypted into 32 vectors in $\mathbb{R}^{(2d+6)}$ and each query vector into 16 matrices in $\mathbb{R}^{(2d+6) \times (2d+6)}$. Thus, AME is space-inefficient.
When calculating each SDC, AME conducts 16 vector-matrix multiplications and 16 inner products between vectors, where each vector is in $\mathbb{R}^{(2d+6)}$ and each matrix in $\mathbb{R}^{(2d+6) \times (2d+6)}$. Hence, each SDC in AME requires $64d^2+416d+676$ multiply-and-accumulate operations. 

\emph{Hence, AME is neither cost-efficient nor space-efficient, especially for high-dimensional vectors.} Our empirical study in Section~\ref{ssec:comp_baseline} also justifies this point.




\section{Distance Comparison Encryption}
\label{sec:secure_dc}

In this section, we present our SDC scheme, called Distance Comparison Encryption~(DCE), which can process distance comparisons securely, exactly, and efficiently.

\begin{figure}[t]
	\centering
	\includegraphics[width=\linewidth]{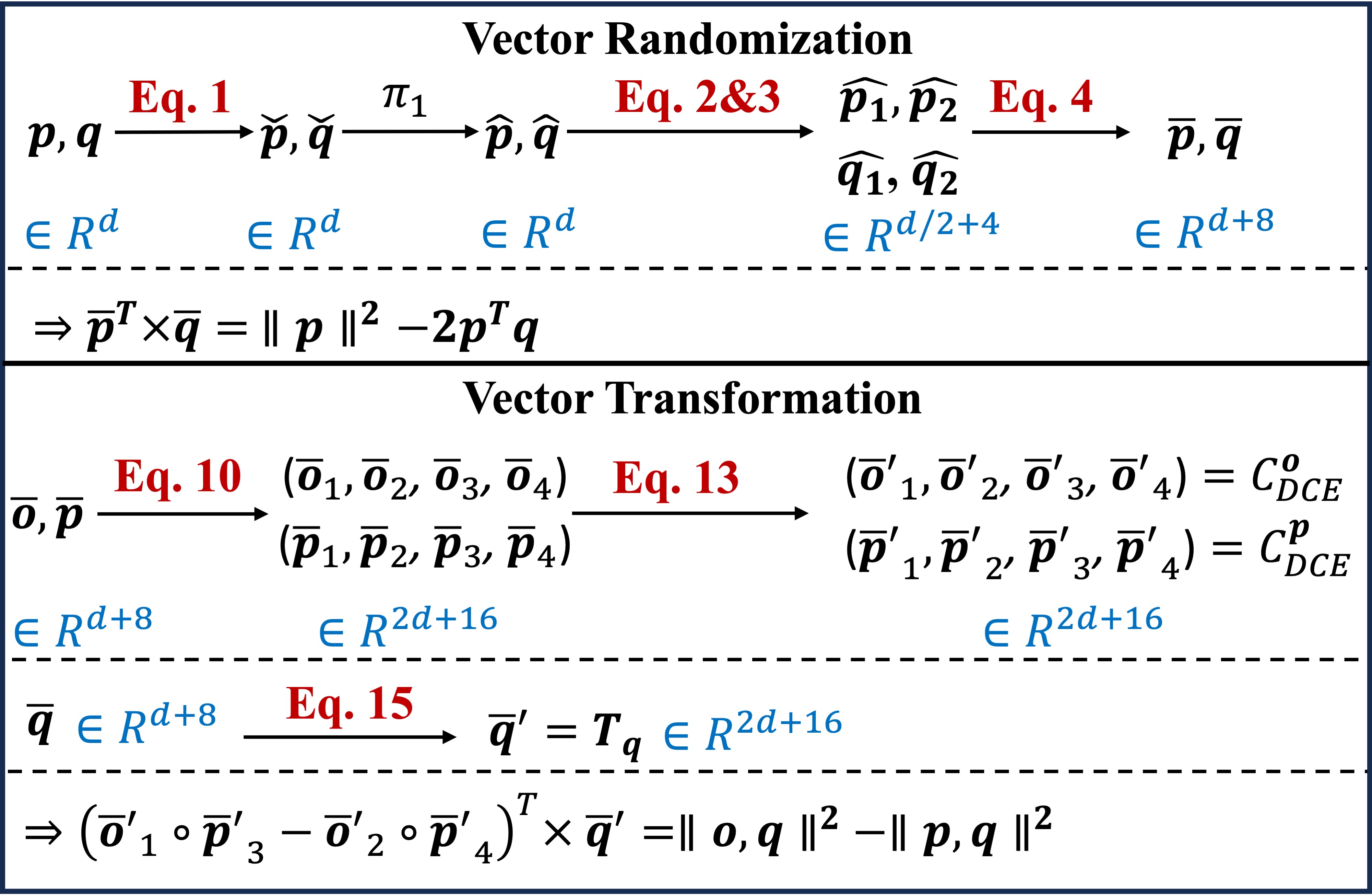}
 \vspace{-3mm}
	\caption{The procedure of our DCE scheme, where $\bm{o, p} \in P$ and $\bm{q}$ is the query vector.}
	\label{fig:dce_roadmap}
 \vspace{-6mm}
\end{figure}

\subsection{Main Idea of Our Scheme}
\label{ssec:dce_idea}

In this part, we present the core idea of DCE scheme, which consists of two phases: (1) \textbf{vector randomization} and (2) \textbf{vector transformation}{, as illustrated in Figure~\ref{fig:dce_roadmap}}. 
The former generates a random vector on top of the original one in order to add randomness and scramble attackers, while the latter further transforms the output of the first phase to produce the final ciphertexts. 

\vspace{1mm}\noindent \textbf{Vector Randomization}. For simplicity, let $\bm{p} = [p_1, \ldots, p_d] \in \mathbb{R}^d$ be a database vector and $\bm{q} = [q_1, \ldots, q_d] \in \mathbb{R}^d$ be a query vector. This phase consists of 4 steps. 

\textit{Step 1:} two new vectors $\check{\bm{p}}$ and $\check{\bm{q}}$ are generated according to the followig equation, with $\check{\bm{p}}^T\check{\bm{q}} = -2\bm{p}^T\bm{q}$:
\begin{equation}
\label{eq:check_pq}
\begin{cases}
\begin{split}
    \check{\bm{p}} &= [p_{1}+p_{2}, p_{1}-p_{2},p_{3}+p_{4}, \cdots, p_{d-1}+p_{d},p_{d-1}-p_{d}]^T, \\ %
    \check{\bm{q}} &=-[q_1+q_2, q_1-q_2, q_3+q_4, \ldots, q_{d-1}+q_d, q_{d-1} - q_d]^T.
\end{split}
\end{cases}
\end{equation}

\textit{Step 2: random permutation.} Let $\pi_1 : \mathbb{R}^d \to \mathbb{R}^d$ be a random permutation on vector. Then, we have $\widehat{\bm{p}}=\pi_{1}(\check{\bm{p}})$ and $\widehat{\bm{q}}=\pi_{1}(\check{\bm{q}})$. Moreover, $\widehat{\bm{p}}^T\widehat{\bm{q}} = \check{\bm{p}}^T\check{\bm{q}} = -2\bm{p}^T\bm{q}$ holds. 

\textit{Step 3: vector spliting and random numbers addition.} First, four random nubmers $r_1, r_2, r_3, r_4 \in \mathbb{R}$ are randomly selected for all database and query vectors. Then, we select five random numbers $\alpha_{\bm{p},1}, \alpha_{\bm{p},2}, r_{\bm{p},1}^{\prime}, r_{\bm{p},2}^{\prime}, r_{\bm{p},3}^{\prime} \in \mathbb{R}$ for $\bm{p} \in P$, and two random numbers $\beta_{\bm{q},1}, \beta_{\bm{q},2} \in \mathbb{R}$ for each query vector $\bm{q}$. Next, we define a variable $ \gamma_{\bm{p}} = (||\bm{p}||^2 - r_{1}^{\prime }r_{1} - r_{2}^{\prime }r_{2}- r_{3}^{\prime }r_{3}) / r_{4}$ accordingly.
$\widehat{\bm{p}}$ is divided into two vectors with random numbers via the following equation: 
\begin{equation}
\label{eq:hat_p}
\begin{cases}
\begin{split}
    \widehat{\bm{p}}_1 &= [\widehat{p}_1, \widehat{p}_2, \ldots, \widehat{p}_{d/2}, \alpha_{\bm{p},1}, -\alpha_{\bm{p},1}, r_{\bm{p},1}^{\prime }, r_{\bm{p},2}^{\prime}]^T, \\
    \widehat{\bm{p}}_2 &= [\widehat{p}_{d/2+1}, \widehat{p}_{d/2+2}, \ldots, \widehat{p}_{d}, \alpha_{\bm{p},2}, \alpha_{\bm{p},2}, r_{\bm{p},3}^{\prime }, \gamma_{\bm{p}}]^T.
\end{split}
\end{cases}
\end{equation}
Similarly, $\widehat{\bm{q}}$ is divided into two vectors:
\begin{equation}
\label{eq:hat_q}
\begin{cases}
\begin{split}
    \widehat{\bm{q}}_1 &= [\widehat{q}_1, \widehat{q}_2, \ldots, \widehat{q}_{d/2}, \beta_{\bm{q},1}, \beta_{\bm{q},1}, r_1, r_2]^T, \\
    \widehat{\bm{q}}_2 &= [\widehat{q}_{d/2+1}, \widehat{q}_{d/2+2}, \ldots, \widehat{q}_{d}, \beta_{\bm{q},2}, -\beta_{\bm{q},2}, r_3, r_4]^T.
\end{split}
\end{cases}
\end{equation}
Based on this, we have $ [\widehat{\bm{p}}_1^T, \widehat{\bm{p}}_2^T] \times [\widehat{\bm{q}}_1^T, \widehat{\bm{q}}_2^T]^T = ||\bm{p}||^2 -2\bm{p}^T\bm{q}$.  

\textit{Step 4: matrix encryption and random permutation.} Let $M_1, M_2 \in \mathbb{R}^{(d/2+4)\times (d/2+4)}$ be two random invertible matries, and $\pi_2 : \mathbb{R}^{(d+8)} \to \mathbb{R}^{(d+8)}$ be a random permutation on vector. We define two variables $\bar{\bm{p}}, \bar{\bm{q}}$ as follows: 
\begin{equation}
\label{eq:bar_pq}
\begin{cases}
\begin{split}
    &\bar{\bm{p}} = \pi_2([\widehat{\bm{p}}_1^TM_1, \widehat{\bm{p}}_2^TM_2]^T), \\
    & \bar{\bm{q}} = \pi_2([(M_1^{-1} \widehat{\bm{q}}_1)^T, (M_2^{-1}\widehat{\bm{q}}_2)^T]^T). \\
\end{split}
\end{cases}
\end{equation}
Further, we have the following equation:
\begin{equation}
\label{eq:random}
    \bar{\bm{p}}^T \bar{\bm{q}} = [\widehat{\bm{p}}_1^T, \widehat{\bm{p}}_2^T] \times [\widehat{\bm{q}}_1^T, \widehat{\bm{q}}_2^T]^T = ||\bm{p}||^2 -2\bm{p}^T\bm{q}.
\end{equation}

According to the four steps of vector randomization, we generate $\bar{\bm{p}}, \bar{\bm{q}} \in \mathbb{R}^{d+8}$ for $\bm{p}, \bm{q} \in \mathbb{R}^d$ respectively. 

\vspace{1mm}\noindent \textbf{Vector Transformation}. 
Let $\bm{o} \in \mathbb{R}^d$ be another database vector and $\bar{\bm{o}} \in \mathbb{R}^{d+8}$ be its result of vector randomization. To answer whether $dist(\bm{o}, \bm{q}) < dist(\bm{p}, \bm{q})$,  
it equal to determine the sign of $\bar{\bm{o}}^T\bar{\bm{q}} - \bar{\bm{p}}^T\bar{\bm{q}}$ according to Equation~\ref{eq:random}. Before further discussion, we define several element-wise operators between two vectors. Let $\bm{a} = [a_1, \ldots, a_d]$ and $\bm{b} = [b_1, \ldots, b_d]$. 
\begin{itemize}
    \item \textit{Element-Wise Addition:} $\bm{a} + \bm{b} = [a_1+b_1, \ldots, a_d+b_d]$;
    \item \textit{Element-Wise Minus:} $\bm{a} - \bm{b} = [a_1 - b_1, \ldots, a_d - b_d]$;
    \item \textit{Element-Wise Multiplication:} $\bm{a} \circ \bm{b} = [a_1 \cdot b_1, \ldots, a_d \cdot b_d]$;
    \item \textit{Element-Wise Division:} $\bm{a} / \bm{b} = [a_1 / b_1, \ldots, a_d / b_d]$.
\end{itemize}
Let $\bm{1}_d$ be the $d$-dimensional vector with all dimensions are $1$, then we have
\begin{equation}
\label{eq:add2multi}
    2\bm{a} + 2\bm{b} = (\bm{a}+\bm{1}_d) \circ (\bm{b}+\bm{1}_d) - (\bm{a}-\bm{1}_d) \circ (\bm{b}-\bm{1}_d).
\end{equation}
Let $\bm{c}, \bm{d} \in \mathbb{R}^d$ be another two vectors, we have 
\begin{equation}
\label{eq:mul_div_exhange}
    (\bm{a} \circ \bm{b}) / (\bm{c} \circ \bm{d}) = (\bm{a} / \bm{c}) \circ (\bm{b} / \bm{d}).
\end{equation}
 
Now, let us focus on how to determine the sign of $\bm{o}^T\bm{q} - \bm{p}^T\bm{q}$ via vector transformation. 
Here, let us introduce a random invertible matrix $M_3 \in \mathbb{R}^{(2d+16) \times (2d+16)}$, which is further divided into two sub-matrices, i.e., $M_{up}, M_{down} \in \mathbb{R}^{(d+8) \times (2d+16)}$. $M_{up}$ represents the first $d+8$ rows of $M_3$ and $M_{down}$ the other $d+8$ rows. Hence, we have 
\begin{equation}
\label{eq:dc0}
\begin{split}
&\bar{\bm{o}}^T\bar{\bm{q}} - \bar{\bm{p}}^T\bar{\bm{q}}
= [\bar{\bm{o}}^T, \bar{\bm{p}}^T] \times [\bar{\bm{q}}^T, -\bar{\bm{q}}^T]^T \\
&= \bigg([\bar{\bm{o}}^T, \bar{\bm{p}}^T] \times \begin{bmatrix} M_{up} \\ M_{down} \end{bmatrix} \bigg) \times \bigg(M_3^{-1} \times [\bar{\bm{q}}^T, -\bar{\bm{q}}^T]^T \bigg).
\end{split}
\end{equation}

Vector transformation is on top of Equation~\ref{eq:dc0}. We further extend the first term\eat{ of RHS} in the second line of Equation~\ref{eq:dc0} and derive the following equation with Equation~\ref{eq:add2multi}:
\begin{equation}
\label{eq:dc1}  
\begin{split}
\mathcal{F}_1(\bar{\bm{o}}, \bar{\bm{p}}) &= 2[\bar{\bm{o}}^T, \bar{\bm{p}}^T] \times \begin{bmatrix} M_{up} \\ M_{down} \end{bmatrix} \\
&= 2 \bar{\bm{o}}^TM_{up} + 2 \bar{\bm{p}}^TM_{down} \\
&= (\bar{\bm{o}}^TM_{up} + \bm{1}_{2d+16}) \circ 
   (\bar{\bm{p}}^TM_{domn} + \bm{1}_{2d+16}) - \\
&  \quad\ (\bar{\bm{o}}^TM_{up} - \bm{1}_{2d+16}) \circ 
   (\bar{\bm{p}}^TM_{domn} - \bm{1}_{2d+16}).
\end{split}
\end{equation}
For simplicity, we denote four vectors for $\bar{\bm{o}}$ and $\bar{\bm{p}}$ as follows:
\begin{equation}
\label{eq:4opvec}
\begin{cases}
\begin{split}
    \bar{\bm{o}}_1 &= \bar{\bm{o}}^TM_{up} + \bm{1}_{2d+16}, 
          &\bar{\bm{p}}_1 &= \bar{\bm{p}}^TM_{up} + \bm{1}_{2d+16} \\
    \bar{\bm{o}}_2 &= \bar{\bm{o}}^TM_{up} - \bm{1}_{2d+16}, 
          &\bar{\bm{p}}_2 &= \bar{\bm{p}}^TM_{up} - \bm{1}_{2d+16} \\
    \bar{\bm{o}}_3 &= \bar{\bm{o}}^TM_{down} + \bm{1}_{2d+16}, 
          &\bar{\bm{p}}_3 &= \bar{\bm{p}}^TM_{down} + \bm{1}_{2d+16} \\
    \bar{\bm{o}}_4 &= \bar{\bm{o}}^TM_{down} - \bm{1}_{2d+16}, 
         &\bar{\bm{p}}_4 &= \bar{\bm{p}}^TM_{down} - \bm{1}_{2d+16}
\end{split}
\end{cases}.
\end{equation}
With those vectors, we can use the following equation:
\begin{equation}
\label{eq:dc2}
\mathcal{F}_1(\bar{\bm{o}}, \bar{\bm{p}}) = \bar{\bm{o}}_1 \circ \bar{\bm{p}}_3 - \bar{\bm{o}}_2 \circ \bar{\bm{p}}_4  .
\end{equation}
Let $k_{v1}, k_{v2}, k_{v3}, k_{v4} \in \mathbb{R}^{2d+16}$ be four random vectors and satisfy $k_{v1} \circ k_{v3} = k_{v2} \circ k_{v4}$. We have the following equation by combining Equation~\ref{eq:mul_div_exhange} and Equation~\ref{eq:dc2}: 
\begin{equation}
\label{eq:f2}
\begin{split}
\mathcal{F}_2(\bar{\bm{o}}, \bar{\bm{p}}) &= \mathcal{F}_1(\bar{\bm{o}}, \bar{\bm{p}}) / (k_{v1} \circ k_{v3}) \\
&= (\bar{\bm{o}}_1 \circ \bar{\bm{p}}_3) / (k_{v1} \circ k_{v3}) - (\bar{\bm{o}}_2 \circ \bar{\bm{p}}_4) / (k_{v2} \circ k_{v4}) \\
&= (\bar{\bm{o}}_1 / k_{v1}) \circ (\bar{\bm{p}}_3 / k_{v3}) - (\bar{\bm{o}}_2 / k_{v2}) \circ (\bar{\bm{p}}_4 / k_{v4}).
\end{split}
\end{equation}
For simplicity, we define another four vectors for $\bar{\bm{o}}$ and $\bar{\bm{p}}$ with two random numbers $r_{\bm{o}}, r_{\bm{p}} \in \mathbb{R}^+$ as follows:
\begin{equation}
\label{eq:4opvec2}
\begin{cases}
    \bar{\bm{o}}_1^{\prime} = r_{\bm{o}} \cdot \bar{\bm{o}}_1 / k_{v1}, \quad \bar{\bm{p}}_1^{\prime} = r_{\bm{p}} \cdot \bar{\bm{p}}_1 / k_{v1} \\
    \bar{\bm{o}}_2^{\prime} = r_{\bm{o}} \cdot \bar{\bm{o}}_2 / k_{v2}, \quad \bar{\bm{p}}_2^{\prime} = r_{\bm{p}} \cdot \bar{\bm{p}}_2 / k_{v2} \\
    \bar{\bm{o}}_3^{\prime} = r_{\bm{o}} \cdot \bar{\bm{o}}_3 / k_{v3}, \quad \bar{\bm{p}}_3^{\prime} = r_{\bm{p}} \cdot \bar{\bm{p}}_3 / k_{v3} \\
    \bar{\bm{o}}_4^{\prime} = r_{\bm{o}} \cdot \bar{\bm{o}}_4 / k_{v4}, \quad \bar{\bm{p}}_4^{\prime} = r_{\bm{p}} \cdot \bar{\bm{p}}_4 / k_{v4}.
\end{cases}
\end{equation}
Here, we introduce the two random numbers to add randomness for the sake of security. By combining Equation~\ref{eq:f2} and \ref{eq:4opvec2}, we have the following equation:
\begin{equation}
\label{eq:dc4}
\mathcal{F}_3(\bar{\bm{o}}, \bar{\bm{p}}) = r_{\bm{o}}r_{\bm{p}}\mathcal{F}_2(\bar{\bm{o}}, \bar{\bm{p}}) = \bar{\bm{o}}_1^{\prime} \circ \bar{\bm{p}}_3^{\prime} - \bar{\bm{o}}_2^{\prime} \circ \bar{\bm{p}}_4^{\prime}.
\end{equation}

Now, let us consider the second term\eat{ of RHS} in the second line of Equation~\ref{eq:dc0}, i.e., $M_3^{-1} \times [\bar{\bm{q}}^T, -\bar{\bm{q}}^T]^T$. With another random number $r_{\bm{q}} \in \mathbb{R}^+$ and the two vectors $k_{v2}, k_{v4}$, we define the following vector for $\bar{\bm{q}}$:
\begin{equation}
\label{eq:vec_q}
\begin{split}
\bar{\bm{q}}^{\prime} = r_{\bm{q}} \cdot M_3^{-1} \times [\bar{\bm{q}}^T, -\bar{\bm{q}}^T]^T \circ (k_{v2} \circ k_{v4}).
\end{split}
\end{equation}
Then, we combine the equations above and have the following equation and prove it in Theorem~\ref{tm:dce_proof}:
\begin{equation}
\label{eq:sdc_idea}
\mathcal{F}_3(\bar{\bm{o}},  \bar{\bm{p}})^T \times \bar{\bm{q}}^{\prime} = 2r_{\bm{o}} r_{\bm{p}} r_{\bm{q}} (||\bm{o}||^2 -2\bm{o}^T\bm{q} - ||\bm{p}||^2 +2\bm{p}^T\bm{q}).
\end{equation}
According to Equation~\ref{eq:sdc_idea}, we can determine whether or not $dist(\bm{o}, \bm{q}) < dist(\bm{p}, \bm{q})$, by precomputing four vectors of each database vector $\bm{p} \in P$, i.e., $\bar{\bm{p}}_1, \bar{\bm{p}}_4, \bar{\bm{p}}_3, \bar{\bm{p}}_4$ as in Equation~\ref{eq:4opvec2}, and $\bar{\bm{q}}^{\prime}$ as in Equation~\ref{eq:vec_q} for a given query $\bm{q}$. 

\subsection{Our DCE Scheme}
\label{ssec:dce}

In this section, we elaborate on our DCE scheme. 
It contains four key functions: (1) \texttt{KeyGen} generates secret keys $SK$, (2) \texttt{Enc} employs $SK$ to encrypt each database vector, (3) \texttt{TrapGen} produces the trapdoor for each query vector with $SK$, and (4) \texttt{DistanceComp} conducts secure distance comparison. As follows, we introduce them in detail, respectively.

\vspace{1mm}\noindent 
(1) \texttt{KeyGen$(1^{\zeta},d)\to SK$.} Given a security parameter $\zeta$ and $d$, it generates $SK=\{M_{1}, M_{2}, M_{3},\pi_{1}, \pi_{2},r_{1},r_{2},\linebreak r_{3},r_{4},k_{v1},k_{v2},k_{v3},k_{v4}\}$. 
As presented in Section~\ref{ssec:dce_idea}, 
the random invertible matrices $M_{1}, M_2 \in \mathbb{R}^{(d/2+4) \times (d/2+4)}$, 
the random permutation functions $\pi_1{:}\mathbb{R}^d\to\mathbb{R}^d,\pi_2{:}\mathbb{R}^{(d+8)}\to\mathbb{R}^{(d+8)}$ and
the four random numbers $r_1, r_2, r_3, r_4 \in \mathbb{R}$ are used in the phase of vector randomization. 
On the other hand, the random invertible matrix $M_3 \in \mathbb{R}^{(2d+16)\times (2d+16)}$ and
four random vectors $k_{v1}, k_{v2}, k_{v3}, k_{v4} \in \mathbb{R}^{(2d+16)}$ are used for vector transformation. Note that $M_3$ is divided into two submatrices that contain the first half rows and the second half rows, respectively. Moreover, $k_{v1}\circ k_{v3}=k_{v2}\circ k_{v4}$.

\vspace{1mm}\noindent 
(2) \texttt{Enc$(\bm{p},SK)\to C_{DCE}^{\bm{p}}$.} Given $SK$ and $\bm{p} \in P$, this function outputs the encrypted vector $C_{DCE}^{\bm{p}}$. 
As discussed in Section~\ref{ssec:dce_idea}, it contains two phases. In the phase of vector randomization, according to steps 1-4, a new vector  $\bar{\bm{p}} \in \mathbb{R}^{d+8}$ is generated for $\bm{p}$. In the second phase of vector transformation that takes $\bar{\bm{p}}$ as input, four vectors  $\bar{\bm{p}}_1^{\prime}, \bar{\bm{p}}_2^{\prime}, \bar{\bm{p}}_3^{\prime}, \bar{\bm{p}}_4^{\prime} \in \mathbb{R}^{2d+16}$ are produced via the following two steps. 
\begin{itemize}
    \item [\romannumeral 1] \textbf{matrix encryption:} generate four vectors $\bar{\bm{p}}_1, \bar{\bm{p}}_2, \bar{\bm{p}}_3, \bar{\bm{p}}_4$ via Equation~\ref{eq:4opvec}. 
    \item [\romannumeral 2] \textbf{adding randomness:} generate four vectors $\bar{\bm{p}}_1^{\prime}, \bar{\bm{p}}_2^{\prime}, \bar{\bm{p}}_3^{\prime}, \bar{\bm{p}}_4^{\prime}$ via Equation~\ref{eq:4opvec2}. 
\end{itemize}
Then, we get the ciphertexts of $\bar{\bm{p}}$, i.e., $C_{DCE}^{\bm{p}} = (\bar{\bm{p}}_1^{\prime}, \bar{\bm{p}}_2^{\prime}, \bar{\bm{p}}_3^{\prime}, \bar{\bm{p}}_4^{\prime})$. 

\vspace{1mm}\noindent 
(3) \texttt{TrapGen$(\bm{q},SK)\to T_{\bm{q}}$}: Given $SK$ and a query vector $\bm{q} \in \mathbb{R}^d$, it outputs the trapdoor $T_{\bm{q}}$ via the following two steps{, which are illustrated in Figure~\ref{fig:dce_roadmap}}. 
\begin{itemize}
    \item [\romannumeral 1] \textbf{vector randomization:} generate a new vector  $\bar{\bm{q}} \in \mathbb{R}^{d+8}$ according to the four steps in the first phase in Section~\ref{ssec:dce_idea}. 
    \item [\romannumeral 2] \textbf{vector transfomration:}  taking $\bar{\bm{q}}$ as input, $\bar{\bm{q}}^{\prime} \in \mathbb{R}^{2d+16}$ is generated via Equation~\ref{eq:vec_q}. Hence, we get $T_{\bm{q}} = \bar{\bm{q}}^{\prime}$. 
\end{itemize}

\vspace{1mm}\noindent 
(4) \texttt{DistanceComp$(C_{DCE}^{\bm{o}},C_{DCE}^{\bm{p}},T_{\bm{q}}) \to [\bar{\bm{o}}_1^{\prime} \circ \bar{\bm{p}}_3^{\prime} - \bar{\bm{o}}_2^{\prime} \circ \bar{\bm{p}}_4^{\prime}]^T \bar{\bm{q}}^{\prime}$.} 
According to Equation~\ref{eq:sdc_idea}, we have $Z_{\bm{o}, \bm{p}, \bm{q}} = \texttt{DistanceComp}(C_{DCE}^{\bm{o}},C_{DCE}^{\bm{p}},T_{\bm{q}}) = 2r_{\bm{o}} r_{\bm{p}} r_{\bm{q}} (||\bm{o}||^2 - 2\bm{o}^T\bm{q} - ||\bm{p}||^2 + 2\bm{p}^T\bm{q})$. 
Since $r_{\bm{o}}, r_{\bm{p}}, r_{\bm{q}} \in \mathbb{R}^+$ are all positive numbers, the sign of $Z_{\bm{o}, \bm{p}, \bm{q}}$ exactly answers the distance comparison. To be specific, if $Z_{\bm{o}, \bm{p}, \bm{q}} < 0$, we have $dist(\bm{o}, \bm{q}) < dist(\bm{p}, \bm{q})$. Otherwise,  $dist(\bm{o}, \bm{q}) \geq dist(\bm{p}, \bm{q})$. 

To be formal, we prove the correctness of the function \texttt{DistanceComp($\cdot$)} in the following theorem.
\begin{theorem}
\label{tm:dce_proof}
Denote $Z_{\bm{o},\bm{p},\bm{q}} = $\texttt{DistanceComp}$(C_{DCE}^{\bm{o}},$\\$C_{DCE}^{\bm{p}},T_{\bm{q}})$, then we have
\begin{equation}
  \begin{cases}
  & Z_{\bm{o},\bm{p},\bm{q}} < 0    \quad \Leftrightarrow \quad    dist(\bm{o}, \bm{q}) < dist(\bm{p}, \bm{q}) \\
  & Z_{\bm{o},\bm{p},\bm{q}} \geq 0 \quad  \Leftrightarrow \quad   dist(\bm{o}, \bm{q}) \geq dist(\bm{p}, \bm{q}).
\end{cases}  
\end{equation}
\end{theorem}
\begin{IEEEproof}
$\texttt{DistanceComp}(C_{DCE}^{\bm{o}},C_{DCE}^{\bm{p}},T_q)$
\begin{align*}
&=[\bar{\bm{o}}_1^{\prime} \circ \bar{\bm{p}}_3^{\prime} - \bar{\bm{o}}_2^{\prime} \circ \bar{\bm{p}}_4^{\prime}] \times \bar{\bm{q}}^{\prime}  \hspace{2.8cm}  (\textrm{by definition})\\
&=r_{\bm{o}}r_{\bm{p}}\mathcal{F}_2(\bar{\bm{o}}, \bar{\bm{p}}) \times \bar{\bm{q}}^{\prime} \hspace{4.1cm}  (\because Eq.~\ref{eq:dc4})  \\
&=r_{\bm{o}}r_{\bm{p}}\mathcal{F}_1(\bar{\bm{o}}, \bar{\bm{p}}) / (k_{v1} \circ k_{v3}) \times \bar{\bm{q}}^{\prime} \hspace{2.3cm} (\because Eq.~\ref{eq:f2})\\
&= 2r_{\bm{o}}r_{\bm{p}}r_{\bm{q}} [\bar{\bm{o}}^T, \bar{\bm{p}}^T] \times M_3 / (k_{v1} \circ k_{v3}) \times \hspace{1.35cm}  (\because Eq.~\ref{eq:dc1}) \\
&\quad\quad M_3^{-1} \times [\bar{\bm{q}}^T, -\bar{\bm{q}}^T]^T \circ (k_{v2} \circ k_{v4}) \hspace{1.65cm}  (\because Eq.~\ref{eq:vec_q}) \\ 
&= 2r_{\bm{o}}r_{\bm{p}}r_{\bm{q}} [\bar{\bm{o}}^T, \bar{\bm{p}}^T] \times [\bar{\bm{q}}^T, -\bar{\bm{q}}^T]^T \ (\because k_{v1} \circ k_{v3} = k_{v2} \circ k_{v4}) \\
&= 2r_{\bm{o}}r_{\bm{p}}r_{\bm{q}} (\bar{\bm{o}}^T\bar{\bm{q}} - \bar{\bm{p}}^T\bar{\bm{q}}) \\
&= 2r_{\bm{o}}r_{\bm{p}}r_{\bm{q}} (||\bm{o}||^2 - 2\bm{o}^T\bm{q} - ||\bm{p}||^2 + 2\bm{p}^T\bm{q})) \hspace{1.3cm}  (\because Eq. \ref{eq:random}) \\
& = 2r_{\bm{o}}r_{\bm{p}}r_{\bm{q}} (dist(\bm{o}, \bm{q}) - dist(\bm{p}, \bm{q})).
\end{align*}

Since $r_{\bm{o}}, r_{\bm{p}}, r_{\bm{q}} \in \mathbb{R}^+$, we have $Z_{\bm{o},\bm{p},\bm{q}} < 0 \Leftrightarrow dist(\bm{o}, \bm{q}) < dist(\bm{p}, \bm{q})$ and $Z_{\bm{o},\bm{p},\bm{q}} \geq 0 \Leftrightarrow dist(\bm{o}, \bm{q}) \geq dist(\bm{p}, \bm{q})$. 
\end{IEEEproof}

\emph{Theorem~\ref{tm:dce_proof} proves that our DCE scheme precisely resolves the distance comparison.}
Besides, \emph{our DCE scheme efficiently handles the distance comparison}, since each SDC computation in DCE only requires $4d+32$ multiply-and-accumulate operations and a time cost of $O(d)$. The dimension of each encrypted database vector is $8d+64$, while that of each encrypted query vector is $2d+16$.
\emph{Furthermore, our DCE scheme is proved to be IND-KPA security in Section~\ref{sec:secana}.}

\eat{
\begin{table}[t]
    \caption{Comparisons of SDC schemes.}
    \label{tb:cmp_sdc}
    \resizebox{\linewidth}{!}{
    \begin{tabular}{|c|c|c|c|}
    \hline
    Methods & $dim_{\bm{p}}$ & $dim_{\bm{q}}$ & $cost_{dc}$ \\
    \hline
    AME & $64d+192$ & $32d+96$ & $64d^2+416d+676$ \\
    \hline
    DCE & $8d+64$ & $2d+16$ &  $4d+32$ \\
    \hline
    \end{tabular}}
\end{table}
}

With our DCE scheme, we can securely conduct a $k$-NN search through a linear scan method utilizing a max heap $H$ to retain the current top $k$ results. Each database vector $\bm{p} \in P$ is considered a candidate in the linear scan, assessing if it outperforms those in $H$ at the complexity of $O(d \times \log k)$. Hence, employing the linear scan method with our DCE scheme incurs a cost of $(n \times d \times \log k)$, which can be prohibitive, particularly for large-scale datasets. As a result, in the next section, we consider designing an efficient index to diminish the count of SDC computations while compromising minimal accuracy.

\section{Our PP-ANNS Scheme}
\label{sec:ours}

\begin{figure}[t]
	\centering
	\includegraphics[width=0.9\linewidth]{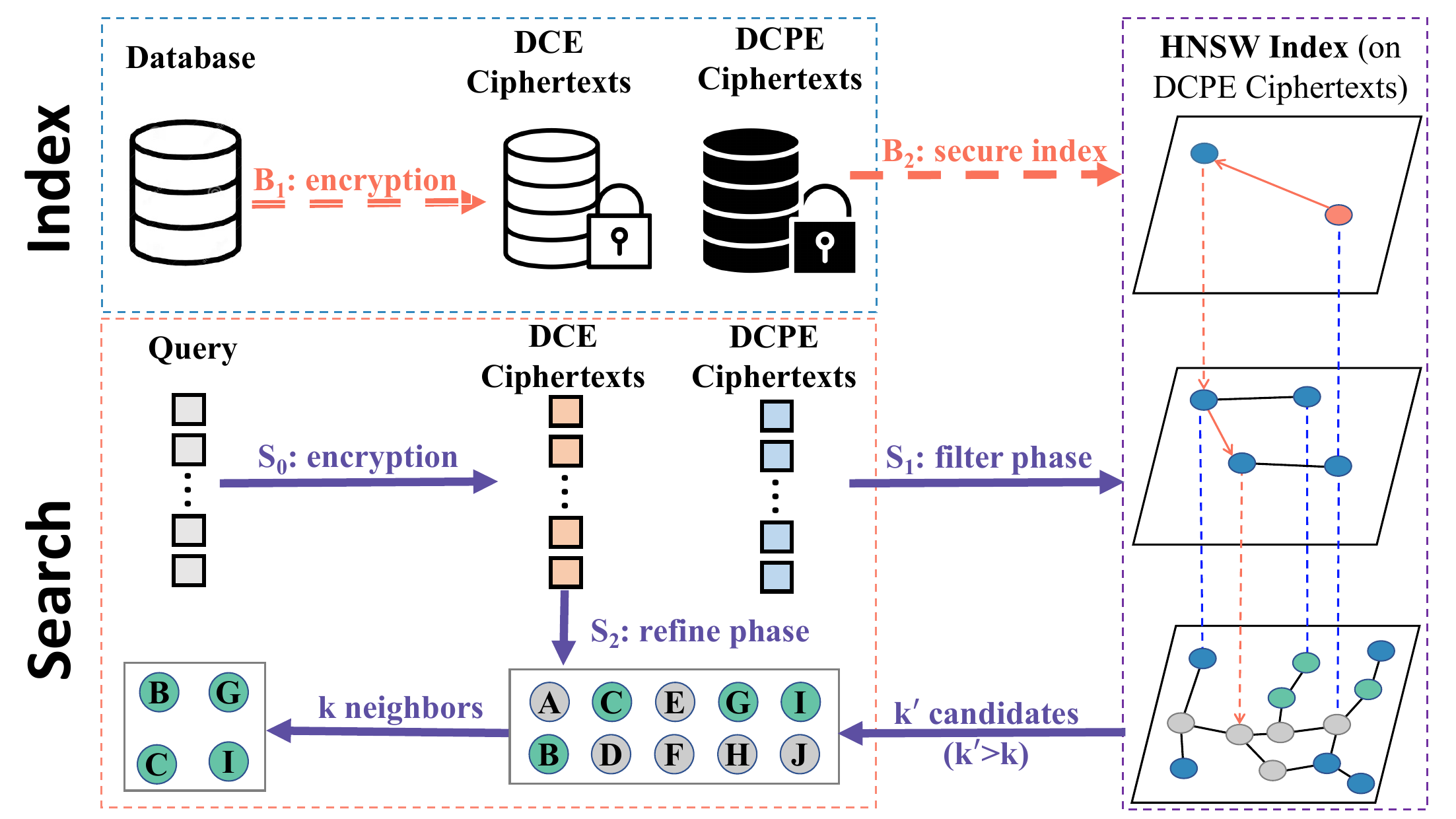}
 \vspace{-3mm}
	\caption{The Overview of our PP-ANNS scheme.
 }
	\label{fig:System Overview}
 \vspace{-6mm}
\end{figure}

In this section, building upon the DCE scheme introduced earlier, we present our whole PP-ANNS scheme depicted in Figure~\ref{fig:System Overview}. Our scheme comprises two components: \textbf{index} and \textbf{search}. We then delve into each of them in detail and analyze the computational and spatial costs of our PP-ANNS scheme.

\subsection{Index Structure}
\label{ssec:index}


To reduce the number of expensive SDC computations of DCE, we propose a privacy-preserving index based on two techniques, i.e., distance-comparison-preserving encryption (DCPE)~\cite{fuchsbauer2022approximate} and HNSW~\cite{malkov2018efficient}. 
The former is a privacy-preserving encryption technique for vectors, while the latter is the state-of-the-art method for $k$-ANNS on vectors. 

\vspace{1mm}
\noindent\textbf{DCPE.}
As mentioned in Section~\ref{ssec:dcpe}, DCPE employs a $\beta$-DCP function to encrypt both database vectors and query vectors. 
Moreover, \cite{fuchsbauer2022approximate} proposes such an instance called Scale-and-Perturb (SAP).  
There are two secret keys: (1) the scaling factor $s \in \mathbb{R}$ that is a random number, and (2) the permutation factor $\beta \in [\sqrt{M}, 2M\sqrt{d}]$, where $M = \max_{\bm{p} \in P} \max_{i=1}^d |p_i|$. 
With those two secret keys, the SAP ciphertexts $C^{\bm{p}}_{SAP}$ for each $\bm{p} \in P$ is computed as in Algorithm~\ref{alg:sap}. 
It first generates a random vector $\bm{u}$ drawn from the Guassian distribution $\mathcal{N}(\bm{0}_d, \bm{I}_d)$ (Line 1), where $\bm{0}_d$ is with all zeros and $\bm{I}_d \in \mathbb{R}^{d \times d}$ is the identity matrix, and a random number $x^{\prime}$ drawn from a uniform distribution $\mathcal{U}(0,1)$ (Line 2). 
Then, we obtain $x$ in Line 3, which will be the norm of the vector $\bm{\lambda}_{\bm{p}}$ (Line 4). Finally, the ciphertext of $\bm{p}$ is the addition of two vectors, i.e., scaled $\bm{p}$ and a random vector $\bm{\lambda}_{\bm{p}}$ that is randomly drawn in the ball $B(\bm{0}_d, s\beta / 4)$. 
Here, we slightly modify the SAP function in Algorithm~\ref{alg:sap}, which does not return the information for decryption. This is because we store $\{C^{\bm{p}}_{SAP} | \bm{p} \in P\}$ on the server and we will not decrypt the ciphertext anywhere for security. 
Moreover, SAP uses $dist(C^{\bm{p}}_{SAP}, C^{\bm{q}}_{SAP})$ as an approximation to $dist(\bm{p}, \bm{q})$.

Notably, $\beta$ is key to the balance between the approximation of encrypted vectors and the data privacy, i.e., smaller $\beta$ leads to a more accurate distance between encrypted vectors, but at the risk of revealing more distance information. 

\begin{algorithm}[t]
\SetKwInOut{Input}{Input}
\SetKwInOut{Output}{Output}
\label{alg:sap}
\caption{$Enc_{SAP} (s, \beta, \bm{p})$}
\Input{secret keys $(s, \beta)$, a vector $\bm{p}$}
\Output{the ciphertext of $\bm{p}$ $C_{SAP}^{\bm{p}}$}
    $\bm{u} \gets \mathcal{N}(\bm{0}_d, \bm{I}_d)$\;
    $x^{\prime} \gets \mathcal{U}(0,1)$\;
    $x \gets \frac{s\beta}{4}(x^{\prime})^{1/d}$\;
    $\bm{\lambda}_{\bm{p}} \gets x \cdot \frac{\bm{u}} {\|\bm{u}\|}$\;
    $C_{SAP}^{\bm{p}} \gets s \cdot \bm{p} + \bm{\lambda}_{\bm{p}}$\;
    \Return $C_{SAP}^{\bm{p}}$\;
\end{algorithm}








\vspace{1mm}
\noindent\textbf{HNSW.} Let $C_{SAP}^{P} = \{ C_{SAP}^{\bm{p}} | \bm{p} \in P\}$ denote the encrypted vectors of $P$. Note that $C_{SAP}^{\bm{p}}$ is still $d$-dimensional. The data owner then builds an HNSW \cite{malkov2018efficient} graph, one of the SOTA $k$-ANNS method~\cite{survey2021,Dpg}, to index $C_{SAP}^{P}$. 
As shown in Figure \ref{fig:System Overview}, HNSW contains multiple layers, where each layer has a graph structure that manages a subset of $C_{SAP}^{P}$. The higher layers contain fewer nodes than the lower layers and layer 0 contains the whole set $C_{SAP}^{P}$. HNSW is built from scratch and inserts each vector $\bm{v} \in C_{SAP}^{P}$ into the graph one by one. 

Notably, data owner does not build the HNSW graph on the plaintext database, since each edge in the graph reflects the neighboring relationship between the corresponding vectors. On the other hand, encrypted DCPE vectors are just the approximation of the original vectors and thus the edges of HNSW built on them do not reflect the exact neighborhood for each vector, which enhances the data privacy. 




In summary, the index in our PP-ANNS scheme contains three parts, i.e., $C_{SAP}^P$, the HNSW index built on $C_{SAP}^P$, and $C_{DCE}^P = \{C_{DCE}^{\bm{p}} | \bm{p} \in P \}$. {Notably, our approach can leverage other proximity graph-based approaches for $k$-ANNS like the navigating spreading-out graph \cite{Nsg} and the $\tau$-monotonic neighborhood graph \cite{taumg} to substitute HNSW for indexing the DCPE-encrypted vectors.}



\subsection{Search Method}
\label{ssec:search}

\begin{algorithm}[t]
\SetKwInOut{Input}{Input}
\SetKwInOut{Output}{Output}
\label{alg:search}
\caption{Search $(T_{\bm{q}}, C_{SAP}^{\bm{q}}, C_{SAP}^P, C_{DCE}^P)$}
\Input{encrypted query vector: $T_{\bm{q}}, C_{SAP}^{\bm{q}}$,encrypted database: $C_{SAP}^P, C_{DCE}^P$}
\Output{the $k$-ANNS result}
    {\color{blue}\tcp{\textbf{\texttt{filter phase:}}}}
    $R'\leftarrow$ the results of $k'$-ANNS on HNSW built on $C_{SAP}^P$ where the query is $C_{SAP}^{\bm{q}}$ \;
    {\color{blue}\tcp{\textbf{\texttt{refine phase:}}}}
    Initialize a max heap $H = \emptyset$\;
    \For{each $p\in R'$}
    {
         \If{$|H| < k$}{
         Insert $\bm{p}$ into $H$\;
         \textbf{continue}\;
         }
         $\bm{o} \leftarrow H.top()$\;
         \If{$\texttt{DistanceComp}(C_{DCE}^{\bm{o}}, C_{DCE}^{\bm{p}},T_{\bm{q}}) > 0$} 
         {
            Insert $\bm{p}$ into $H$\;
         }
    }
    \Return the elements in $H$\; 
\end{algorithm}

With our index, we now propose a new search method for the $k$-ANNS as shown in Algorithm~\ref{alg:search}. Our search method follows the \emph{filter-and-refine} strategy. In the \emph{filter} phase (Line 1), we conduct $k^{\prime}$-ANNS ($k^{\prime} > k$) on HNSW built on $C_{SAP}^{P}$, which strictly follows the same procedure as the HNSW search method in \cite{malkov2018efficient} and thus returns a set $R^{\prime}$ of high-quality candidates. Note that we verify all the candidates by approximate distances of their SAP encrypted vectors to the encrypted vector $C_{SAP}^{\bm{q}}$ in this phase. In the \emph{refine} phase (Lines 2-9), a max heap $H$ is initialized as empty in Line 2, which is used to store the currently best $k$ neighbors. Then, we try to insert each $\bm{p} \in R^{\prime}$ into $H$. 

{
Note that $k^{\prime}$ is key to the search performance of our method. A larger $k^{\prime}$ increases the probability of finding more accurate results in the \emph{refine} phase, but at the expense of more SDC computations, each of which is $O(d^2)$. 
Hence, $k^{\prime}$ is key to the balance between the accuracy and efficiency of our method, which is investigated in the experiments of Section~\ref{ssec:parasel}. In practice, we employ the grid search method to select the best value of $k^{\prime}$. 
}

\noindent\textbf{Time Complexity.}
We use the \texttt{DistanceComp} function to compare two vectors and determine which one is closer to the query. Hence, each insertion in $H$ may cause at most $\log k$ distance comparisons, each of which costs $O(d)$. As a result, the cost of the \emph{refine} phase is $O(k^{\prime} \cdot d \cdot \log k)$.

\subsection{Cost Analysis}

In this section, we analyze the cost complexity of our approach, including computational cost and space requirement. For the computational cost,
we consider (1) the server-side computational cost, (2) the user-side computational cost, and (3) the communication overhead between the user and server. 

\vspace{1mm}
\noindent 
\textbf{Server-Side Computational Cost.} Search on server consists of two phases, i.e., (1) the \emph{filter} phase that conducts $k^{\prime}$-ANNS on HNSW and (2) the \emph{refine} phase that finds the best $k$ neighbors via SDC computations of DCE. 
According to HNSW \cite{malkov2018efficient}, the complexity of the \emph{filter} phase is $O(d \cdot \log n)$. In \emph{refine} phase, we employ a max heap with at most $k$ elements to obtain the best $k$ neighbors from the $k^{\prime}$ candidates. When dealing with each candidate, it may update the heap with $O(\log k)$ SDC computations of DCE, each of which costs $O(d)$. Hence, the \emph{refine} phase costs $O(d \cdot k^{\prime} \log k)$. In total, the complexity of the search cost on the server is $O(d \cdot (\log n + k^{\prime} \log k))$ for each query.  

\noindent 
\textbf{User-Side Computational Cost.} The user only generates $T_{\bm{q}}$ for each given $\bm{q}$, which contains two phases, i.e., vector randomization and vector transformation. The main cost of the first phase comes from Equation~\ref{eq:bar_pq}, whose complexity is $O(d^2)$, while the cost of the second phase is $O(d^2)$ according to Equation~\ref{eq:vec_q}. Besides, we also compute the DCPE encrypted query vector $C_{SAP}^{\bm{q}}$, whose cost is only $O(d)$. As a result, the user-side time complexity is $O(d^2)$. Hence, the user only performs very slight computation. 

\noindent 
\textbf{Communication Overhead.} During the search procedure, only two messages are needed to transfer between user and server, i.e., (1) user sends the encrypted query vectors (i.e., $C_{SAP}^{\bm{q}}$ and $C_{DCE}^{\bm{q}}$) and the parameter $k$ to the server, which occupies $(36 \cdot d+260)$ bytes in total, and (2) server returns $k$ IDs of found neighbors to the user, which requires only $4 \cdot k$ bytes. As a result, there exists extremely limited communication overhead between the user and the server. 

\noindent 
\textbf{Space Complexity.}
We analyze the space requirement of our method in the server. It contains three parts: (1) the DCPE ciphertexts $C_{SAP}^{P}$, (2) the HNSW index, and (3) the ciphertexts $C_{DCE}^{P}$. The space requirement of $C_{SAP}^{P}$ is the same as $P$, while that of $C_{DCE}^{P}$ is $(8+64/d)$ times that of $P$. As to HNSW, its space requirement is $O(n \cdot m)$, where $m$ is the upper bound of the out-degree for each node in  HNSW. Hence, the space requirement is $O(n \cdot (d+m))$.

{
\subsection{Discussion on Index Maintenance}
\label{ssec:maintenance}

In real-world applications, the vector database needs to support updates, e.g., the addition of new vectors or removal of existing ones. Enhancing the index maintenance significantly increases the practical usability of our method. For insertion, let $\bm{u}$ be the vector to be inserted. The data provider first generates two encrypted vectors, i.e., $C_{SAP}^{\bm{u}}$ and $C_{DCE}^{\bm{u}}$ and then sends them to the server. Like inserting a new point in the original HNSW, the server conducts $k$-ANNS for $C_{SAP}^{\bm{u}}$ and selects diverse neighbors from the returned neighbors, followed by building edges between $\bm{u}$ and selected neighbors. Hence, the HNSW graph is updated for inserting $\bm{u}$. 

As to deleting an existing vector $\bm{u}$, only its in-neighbors will be affected, while its out-neighbors will not. For each in-neighbor $\bm{v}$ of $\bm{u}$, we can reinsert it into HNSW, by finding its $k$-ANN in the current graph and then building edges between $\bm{v}$ and a selected subset of its $k$-ANN returned. Besides, the encrypted vectors of $\bm{u}$, i.e., $C_{SAP}^{\bm{u}}$ and $C_{DCE}^{\bm{u}}$, will be deleted. Unlike the insertion that needs the participation of the data provider, the deletion could be finished solely by the server. 
}
\section{SECURITY ANALYSIS}\label{sec:secana}

In this section, we analyze the security of our DCE scheme and PP-ANNS scheme, respectively.

\subsection{Security of DCE Scheme}
\label{ssec:security_dce}


Our DCE scheme is an encryption that supports searchability. Like AME~\cite{zheng2024achieving}, we prove that our DCE scheme is \eat{selectively}IND-KPA secure in the real/ideal world security model. The real world is our proposal and the ideal world is an ideal function that only leaks some information specified by a leakage function, which is only related to some public information in our proposal. We prove that the real world is indistinguishable from the ideal world. To ensure fairness, we follow the proof steps in \cite{zheng2024achieving}. We first define the leakage function to introduce the real and ideal experiments.

Let $\bm{o}, \bm{p} \in P$ be any two database vectors, and $q$ be an arbitrary query vector. For simplicity, let $C_{\bm{o}}$, $C_{\bm{p}}$ and $T_{\bm{q}}$ be their ciphertexts in our DCE scheme, respectively. Then, the leakage of our DCE scheme is the comparison result between $dist(\bm{o}, \bm{q})$ and $dist(\bm{p}, \bm{q})$, i.e., $\mathcal{L}(\bm{o}, \bm{p}, \bm{q})=$ \texttt{DistanceComp}$(C_{\bm{o}}$, $C_{\bm{p}}, T_{\bm{q}}).$ Based on this leakage function, we define the real and ideal experiments in the following.

\noindent\textbf{Real experiment}. The real experiment involves two participants, i.e., a probabilistic polynomial time adversary $\mathcal{A}$ and a challenger, which interact with each other as follows.

\noindent \textbf{Setup}. In the setup phase, $\mathcal{A}$ first chooses a set $P_1$ that contains $n_1$ $d$-dimensional database vectors and sends them to the challenger.
On receiving $P_1$, the challenger calls \texttt{KeyGen}$(1^\zeta,d)$ to generate a secret key
$SK=\{M_{1},M_{2},M_{3},\pi_{1},\pi_{2},r_1,r_{2},r_{3},r_{4},k_{v1},k_{v2},k_{v3},k_{v4}\}.$ Then the challenger calls \texttt{Enc}$(\bm{p},SK)$ to encrypt vectors $\bm{p} \in P_1$. 

\noindent \textbf{Query phase 1}. In this phase, $\mathcal{A}$ needs to choose a set $Q_1$ with $m_1$ $d$-dimensional query vectors, where $m_1$ is a polynomial number. 
Then, $\mathcal{A}$ sends $Q_1$
to the challenger. On receiving them, the challenger calls \texttt{TrapGen}$(\bm{q},SK)$ to encrypt $\bm{q}$ into $T_{\bm{q}}$ for each query $\bm{q} \in Q_1$. Then, the challenger returns $\left\{T_{\bm{q}} | \bm{q} \in Q_1\right\}$ to $\mathcal{A}$.
 
\noindent \textbf{Challenge phase}. In the challenge phase, the challenger returns $\left\{C_{\bm{p}} | \bm{p} \in P_1 \right\}$ to $\mathcal{A}$. 

\noindent \textbf{Query phase 2}. In this phase, $\mathcal{A}$ needs to choose another set $Q_2$ with $m_2$ $d$-dimensional query vectors and gets the ciphertexts $\left\{T_{\bm{q}} | \bm{q} \in Q_2\right\}$ from the challenger in the same way as {Query phase 1}.

\noindent \textbf{Ideal experiment}. The ideal experiment involves two participants, i.e., a probabilistic polynomial time adversary $\mathcal{A}$ and a simulator with the leakage $\mathcal{L}$, interacting as follows.

\eat{\resizebox{6pt}{!}{\raisebox{-1pt}\textbullet}}\noindent \textbf{Setup}. $\mathcal{A}$ chooses $n_1$ $d$-dimensional database vector set $P_1$ and sends them to the simulator. 
On receiving them, the simulator generates a $(2d+16) \times (2d+16)$ matrix $M_{sim}$ and a collection of random vectors $C_{\bm{p},sim}=(\bar{\bm{p}}_{1,sim}^{\prime}, \bar{\bm{p}}_{2,sim}^{\prime}, \bar{\bm{p}}_{3,sim}^{\prime}, \bar{\bm{p}}_{4,sim}^{\prime})$ as ciphertexts for each $\bm{p} \in P_1$, which satisfy $\bar{\bm{p}}_{1,sim}^{\prime}\circ \bar{\bm{p}}_{3,sim}^{\prime} - \bar{\bm{p}}_{2,sim}^{\prime} \circ \bar{\bm{p}}_{4,sim}^{\prime}= {M^{-1}_{sim}} \times [\bm{v}_r^T, -\bm{v}_r^T]^T$. And $\bm{v}_r$ is a $(d+8)$-dimensional random vector.

\eat{\resizebox{6pt}{!}{\raisebox{-1pt}\textbullet}}\noindent \textbf{Query phase 1}. $\mathcal{A}$ chooses a query set $Q_1$ with $m_1$ query vectors, where $m_1$ is a polynomial number. Then, $\mathcal{A}$ sends $Q_1$ to the simulator. Upon receiving these query vectors, the simulator uses the leakage function $\mathcal{L}$ and $\left\{C_{\bm{p},sim} | \bm{p} \in P_1 \right\}$ to generate the ciphertexts for $Q_1$. For each $\bm{q} \in Q_1$, the simulator generates a ciphertext $T_{\bm{q},sim}$ as follows. 
\begin{itemize}
\item First, the simulator generates two random $n_1\times n_1$ matrices $R_{\bm{q}}$ and $R_{\bm{q}}^\prime$, where $r_{i,j}\in R_{\bm{q}}$
and $r_{i,j}^{\prime}\in R_{\bm{q}}^{\prime}$ are random numbers and satisfy that for $1 \leq i,j \leq n_1$, 
$$\begin{cases}
r_{i,j}>0,r_{i,j}^{\prime} > 0 & \mathcal{L}\big(\bm{o},\bm{p},\bm{q}\big) > 0\\
r_{i,j} \leq 0,r_{i,j}^{\prime} \leq 0 & \mathcal{L}\big(\bm{o},\bm{p},\bm{q}\big) \leq 0\\
\end{cases}.$$
Here, $\bm{o}$, $\bm{p}$ are the $i$-th and $j$-th point in $P_1$, respectively.

\item Second, the simulator randomly chooses a vector $T_{\bm{q},sim}$ such that $T_{\bm{q},sim}={M^T_{sim}}\times [\bm{v}_{rq}, -\bm{v}_{rq}], $
$$\left( \bar{\bm{o}}_{1,sim}^{\prime} \circ \bar{\bm{p}}_{3,sim}^{\prime} - \bar{\bm{o}}_{2,sim}^{\prime} \circ \bar{\bm{p}}_{4,sim}^{\prime} \right)^T \times {T_{\bm{q},sim}}=r_{i,j}, $$  
$$\left( \bar{\bm{p}}_{1,sim}^{\prime} \circ \bar{\bm{o}}_{3,sim}^{\prime} - \bar{\bm{p}}_{2,sim}^{\prime} \circ \bar{\bm{o}}_{4,sim}^{\prime} \right)^T \times {T_{\bm{q},sim}}=r_{i,j}^{\prime}.$$ 
The $\bm{v}_{rq}$ is a random $(d+8)$-dimensional vector. The simulator returns $\left\{T_{\bm{q},sim} | \bm{q} \in Q_1 \right\}$ to $\mathcal{A}$.

\end{itemize}

\eat{\resizebox{6pt}{!}{\raisebox{-1pt}\textbullet}}\noindent \textbf{Challenge phase}. In the challenge phase, the simulator returns $\left\{C_{\bm{p},sim} | \bm{p} \in P_1 \right\}$ to $\mathcal{A}$.

\eat{\resizebox{6pt}{!}{\raisebox{-1pt}\textbullet}}\noindent \textbf{Query phase 2}. $\mathcal{A}$ chooses a query set $Q_2$ with $m_2$ $d$-dimensional query vectors and gets their ciphertexts in the same way as {Query phase 1}.

In the real experiment, the view of $\mathcal{A}$ is $\mathrm{View}_{\mathcal{A},\mathrm{Real}}=\{\{C_{\bm{p}} | \bm{p} \in P_1\}, \{T_{\bm{q}} | \bm{q} \in Q_1 \cup Q_2\}\}$. In the ideal experiment, the view of $\mathcal{A}$ is $\mathrm{View}_{\mathcal{A},\mathrm{Ideal}}=\{\{C_{\bm{p},sim} | \bm{p} \in P_1\},\{T_{\bm{q},sim} | \bm{q} \in Q_1 \cup Q_2\}$. Then, based on the views of $\mathcal{A}$ in the real and ideal experiments, we define the security of the DCE scheme.

\eat{
\begin{definition}
(\textbf{Security of DCE scheme}) DCE scheme is selectively secure with the leakage $\mathcal{L}$ iff for any $\mathcal{A}$ issuing a polynomial number of database vectors encryption and query vectors encryption, there exists a simulator such that the advantage that $\mathcal{A}$ can distinguish the views of real and ideal experiments is negligible, i.e., $|\mathrm{Pr}[\mathrm{View}_{\mathcal{A},\mathrm{Real}}=1]-\mathrm{Pr}[\mathrm{View}_{\mathcal{A},\mathrm{Ideal}}=1]|$ is negligible.
\end{definition}
}
In the following, we prove that the DCE scheme is \eat{selectively}IND-KPA secure with leakage $\mathcal{L}$.


\begin{theorem}
The DCE scheme is \eat{selectively}IND-KPA secure with $\mathcal{L}.$
\end{theorem}
\begin{IEEEproof}
We prove the security of our DCE scheme by proving that $\mathcal{A}$ cannot distinguish the views of real and ideal experiments. 
Since all ciphertexts $\{C_{\bm{p},sim} | \bm{p} \in P_1\}$ and $\{T_{\bm{q},sim} | \bm{q} \in Q_1 \cup Q_2\}$ in the ideal experiment are randomly generated by the simulator, distinguishing the real view and the ideal view is equivalent to distinguish View$_{\mathcal{A},\mathrm{Real}}$ from random ciphertexts. To prove the indistinguishability, we consider three cases:
\begin{itemize}
    \item \textbf{Case 1:} The ciphertexts $\{C_{\bm{p}} | \bm{p} \in P_1\}$ are indistinguishable from random ciphertexts;
    \item \textbf{Case 2:} The ciphertexts $\{T_{\bm{q}} | \bm{q} \in Q_1 \cup Q_2\}\}$ are indistinguishable from random ciphertexts;
    \item \textbf{Case 3:} Each intermediate result $Z_{\bm{o},\bm{p},\bm{q}}$ (as defined in Theorm~\ref{tm:dce_proof}) is indistinguishable from random numbers, where $\bm{o}, \bm{p} \in P_1$.
\end{itemize}

Due to the definition of View$_{\mathcal{A}, \mathrm{Real}}$, we naturally need to consider the indistinguishability of {Case 1} and {Case 2}. Besides, we consider the indistinguishability between the intermediate results and random ciphertexts. In our DCE scheme, $\{ Z_{\bm{o},\bm{p},\bm{q}} | \bm{o}, \bm{p} \in P_1\}$ are the intermediate results with the least number of unknown variables. This is because $Z_{\bm{o},\bm{p},\bm{q}} = 2r_{\bm{o}}r_{\bm{p}}r_{\bm{q}} (dist(\bm{o}, \bm{q}) - dist(\bm{p}, \bm{q}))$ as proved in Theorem~\ref{tm:dce_proof}. Obviously, it has eliminated the unknown matrices in the secret keys. Meanwhile, it is easy to verify that other intermediate results contain more unknown variables than $\{Z_{\bm{o},\bm{p},\bm{q}} | \bm{o}, \bm{p} \in P_1\}$. Thus, if $\{Z_{\bm{o},\bm{p},\bm{q}} | \bm{o}, \bm{p} \in P_1\}$ are indistinguishable from random ciphertexts, all other intermediate results are indistinguishable from random ciphertexts. Thus, we consider {Case 1}, {Case 2}, and {Case 3} in the indistinguishability proof, as described below.

{The ciphertexts} $\{C_{\bm{p}} | \bm{p} \in P_1\}$ {are indistinguishable from random ciphertexts.} In our scheme, $C_{\bm{p}} = (\bar{\bm{p}}_1^{\prime}, \bar{\bm{p}}_2^{\prime}, \bar{\bm{p}}_3^{\prime}, \bar{\bm{p}}_4^{\prime})$. Accoring to Equation~\ref{eq:4opvec2}, \ref{eq:4opvec} and \ref{eq:bar_pq}, we have
\begin{equation}
\label{eq:4pvec}
\begin{cases}
    \bar{\bm{p}}_1^{\prime} = r_{\bm{p}} (\pi_2([\widehat{p}_1^T, \widehat{p}_2^T]^T) \times M_{up} + \bm{1}_{2d+16}) / k_{v1} \\
    \bar{\bm{p}}_2^{\prime} = r_{\bm{p}} (\pi_2([\widehat{p}_1^T, \widehat{p}_2^T]^T) \times M_{down} + \bm{1}_{2d+16}) / k_{v2} \\
    \bar{\bm{p}}_3^{\prime} = r_{\bm{p}} (\pi_2([\widehat{p}_1^T, \widehat{p}_2^T]^T) \times M_{up} - \bm{1}_{2d+16}) / k_{v3} \\
    \bar{\bm{p}}_4^{\prime} = r_{\bm{p}} (\pi_2([\widehat{p}_1^T, \widehat{p}_2^T]^T) \times M_{down} - \bm{1}_{2d+16}) / k_{v4}.
\end{cases}
\end{equation}
Here, $\widehat{p}_1$ and $\widehat{p}_1$ could be derived via Equations \ref{eq:hat_p} and \ref{eq:check_pq}. 


To generate $C_{\bm{p}}$, we add many random numbers such as $r_{\bm{p}}, \alpha_{\bm{p},1}, \alpha_{\bm{p},2}, r^{\prime}_{\bm{p},1}$, $r^{\prime}_{\bm{p},2}$ and $r^{\prime}_{\bm{p},3}$. Then, we add multiply those random numbers by the vector, which makes the results indistinguishable. Besides, we use two random permutations (i.e., $\pi_1, \pi_2$) and four random vectors (i.e., $k_{v1}, k_{v2}, k_{v3}, k_{v4}$) to perturb the specific values in $C_{\bm{p}}$. Thus, these random numbers and the random permutations can guarantee that $\{C_{\bm{p}} | \bm{p} \in P_1\}$ are indistinguishable from random ciphertexts.

{The ciphertexts} $\{T_{\bm{q}} | \bm{q} \in Q_1 \cup Q_2\}\}$ {are indistinguishable from random ciphertexts}. In our scheme, $T_{\bm{q}} = r_q  M_3^{-1} \times [\bar{\bm{q}}^T, -\bar{\bm{q}}^T]^T \circ (k_{v2} \circ k_{v4})$ according to Equation~\ref{eq:vec_q}, where $\bar{\bm{q}}$ could be derived via Equation~\ref{eq:bar_pq}, \ref{eq:hat_q} and \ref{eq:check_pq}. 

To compute $T_{\bm{q}}$, we introduce many random numbers such as $r_{\bm{q}}, \beta_{\bm{q,1}}$ and $\beta_{\bm{q,2}}$. Besides, we use two random permutations (i.e., $\pi_1$ and $\pi_2$) and two random vectors (i.e., $k_{v2}$ and $k_{v4}$) to perturb the specific values in $T_{\bm{q}}$, which make each $T_{\bm{q}}$ for $\bm{q} \in Q_1 \cup Q_2$ indistinguishable from random ciphertexts.

{The intermediate results} $\{ Z_{\bm{o},\bm{p},\bm{q}} | \bm{o}, \bm{p} \in P_1 \}$ {are indistinguishable from random numbers.} As defined in Theorm~\ref{tm:dce_proof}, $Z_{\bm{o},\bm{p},\bm{q}} = 2r_{\bm{o}}r_{\bm{p}}r_{\bm{q}} (dist(\bm{o}, \bm{q}) - dist(\bm{p}, \bm{q}))$. We can see that there still exist three random numbers, i.e., $r_{\bm{o}}, r_{\bm{p}}, r_{\bm{q}}$, which makes $Z_{\bm{o},\bm{p},\bm{q}}$ indistinguishable from random ciphertexts.


In conclusion, we can deduce that View$_{\mathcal{A},\mathrm{Real}}$ is indistinguishable from random ciphertexts. That is, View$_{\mathcal{A},\mathrm{Real}}$ is indistinguishable from View$_{\mathcal{A},\mathrm{Ideal}}$, and $\mathcal{A}$ cannot distinguish the views of real and ideal experiments. Therefore, our DCE scheme is \eat{selectively}IND-KPA secure with the leakage $\mathcal{L}$.
\end{IEEEproof}

\subsection{Security of Our PP-ANNS Scheme}

In this part, we analyze the security of our PP-ANNS scheme. Since we focus on privacy-preserving properties, we show that both database and query requests are privacy-preserving as follows. 

\noindent\eat{\resizebox{6pt}{!}{\raisebox{-1pt}\textbullet}}\textbf{{Database is privacy-preserving.}} In our scheme, the cloud server is semi-honest, so it is curious about the plaintexts of the database vectors. However, all database vectors in the cloud server have been encrypted by DCE and DCPE. The security of DCPE scheme has been proven in \cite{fuchsbauer2022approximate}, while we prove the security of DCE scheme in the last section. 
Moreover, there is no relation between DCE and DCPE schemes, and the cloud server cannot recover plaintexts of database vectors even after obtaining both of them. As a result, nothing is revealed except for the HNSW index. Notably, it contains only approximate neighboring relationships between encrypted vectors. 

\noindent\eat{\resizebox{6pt}{!}{\raisebox{-1pt}\textbullet}}\textbf{{The query vector is privacy-preserving.}} The query vector should be kept secret from the cloud server. Each query vector $\bm{q}$ has been encrypted by both DCE and DCPE schemes before sending to the cloud, and the security of DCE and DCPE can guarantee that the cloud server cannot recover the plaintext of $\bm{q}$ from $T_{\bm{q}}$. Thus, each query vector is privacy-preserving.


\begin{table}[t]
	\centering
	\caption{Statistics of Datasets}
 \vspace{-3mm}
	\label{tab:dataset}
	\begin{tabular}{|c|r|r|r|r|}
            \hline
		\textbf{Datasets} & \textbf{Sift1M} & \textbf{Gist} & \textbf{Glove} & \textbf{Deep1M} \\
		\hline
		\#dimensions & 128 & 960 & 100 & 96 \\
            \hline
		  \#vectors & 1,000,000 & 1,000,000 & 1,183,514 & 1,000,000 \\
            \hline
		\#queries & 10,000 & 1,000 & 10,000 & 10,000 \\
		\hline
	\end{tabular}
 \vspace{-3mm}
\end{table}

\section{Experiments}\label{sec:exp}

In this section, we evaluate the performance of our PP-ANNS method and compare it with other baseline methods to demonstrate its superiority. First of all, we introduce the experimental settings.

\noindent {\textbf{Datasets.}} We conduct experiments on four widely-used datasets, i.e., \texttt{Sift1M}\footnote{\label{fn:sift}{http://corpus-texmex.irisa.fr/}}, \texttt{Gist}\footref{fn:sift}, \texttt{Glove}\footnote{https://nlp.stanford.edu/projects/glove/} and \texttt{Deep1M}\footnote{http://sites.skoltech.ru/compvision/noimi/}.The statistics of those datasets have been presented in \cref{tab:dataset}. In addition, we conducted experiments on random samples of the large-scale data Sift1B\footref{fn:sift} with 1 billion SIFT vectors and Deep1B\footnote{https://disk.yandex.ru/d/11eDCm7Dsn9GA/}, in order to verify the scalability of our method. 





\noindent {\textbf{Performance Metrics.}} Our solution is mainly performed on the server, so we focus on the server-side search performance in both efficiency and accuracy. Efficiency is measured by queries per second (QPS), while accuracy is estimated by recall denoted as $Recall@k$. Given a query $q$, let $N^*(q)$ be the exact $k$-nearest neighbors of $q$, while $N(q)$ be the set of $k$ approximate neighbors. We have $Recall@k (q) = |N^* (q) \cap N(q)| / k$, and use the average recalls over the queries to report $Recall@k$. We set the number $k$ to $10$ by default unless specified setting. All results are averaged over 5 runs. 

\begin{figure}[t]
	\centering
\includegraphics[width=0.24\linewidth]{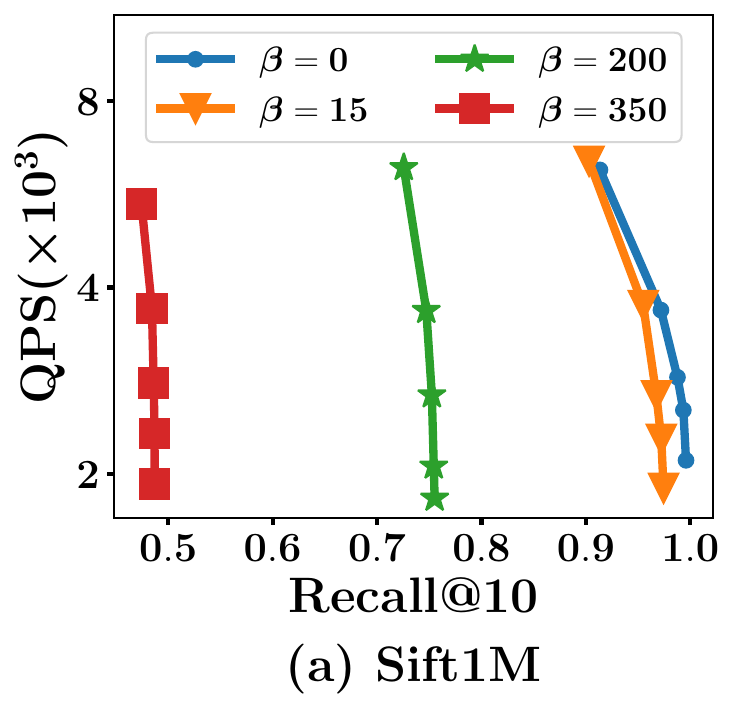}
\includegraphics[width=0.24\linewidth]{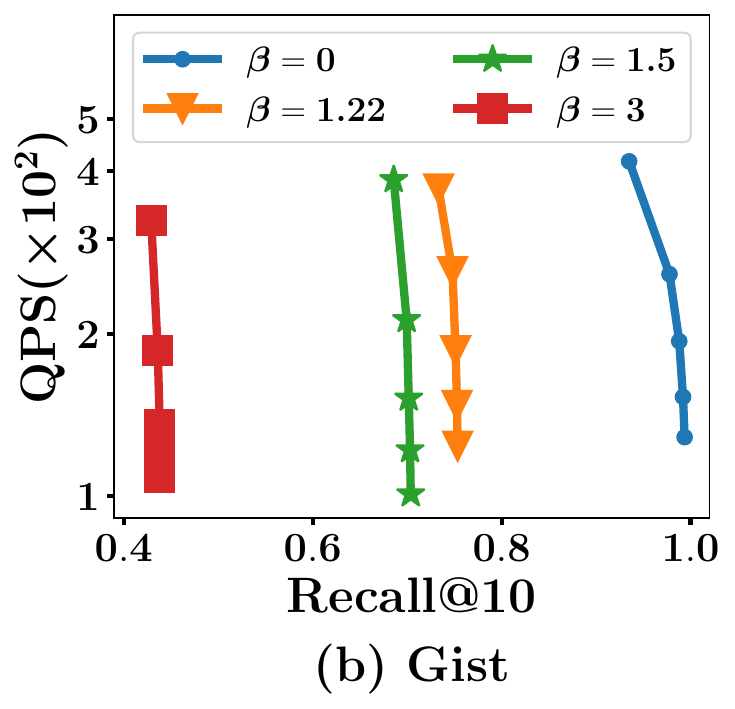}
\includegraphics[width=0.24\linewidth]{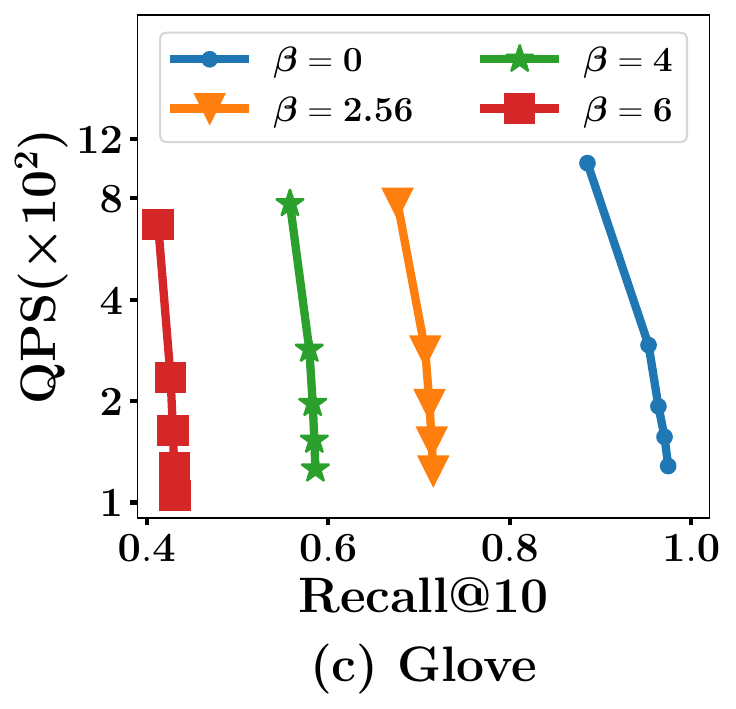}
\includegraphics[width=0.24\linewidth]{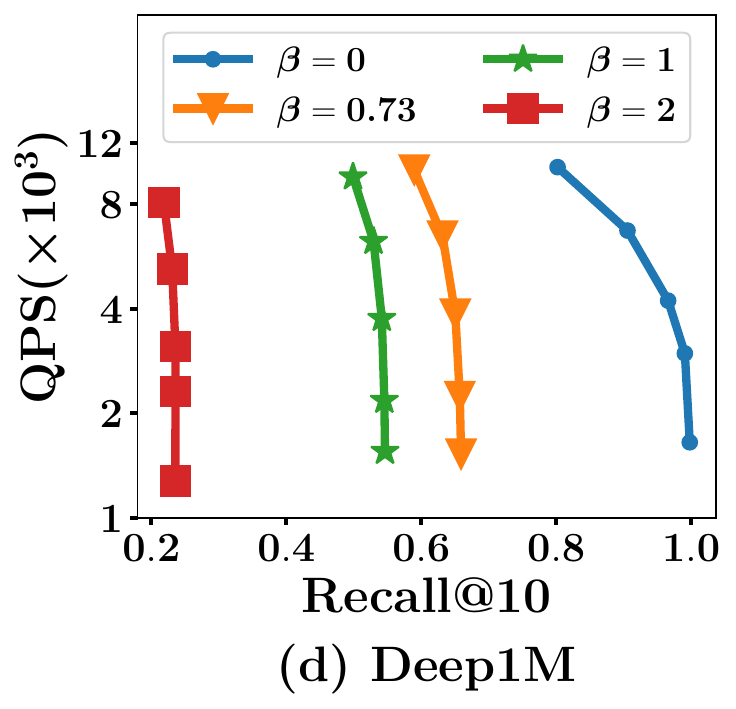}
	\caption{The effects of $\beta$ on the search performance in the \emph{filter} phase. {Note that $\beta = 0$ indicates that there exists no noise in the DCPE encrypted vectors. }}
	\label{fig:beta_recall}
 \vspace{-6mm}
\end{figure}

\begin{figure}[t]
	\centering
    \includegraphics[width=0.8\linewidth]{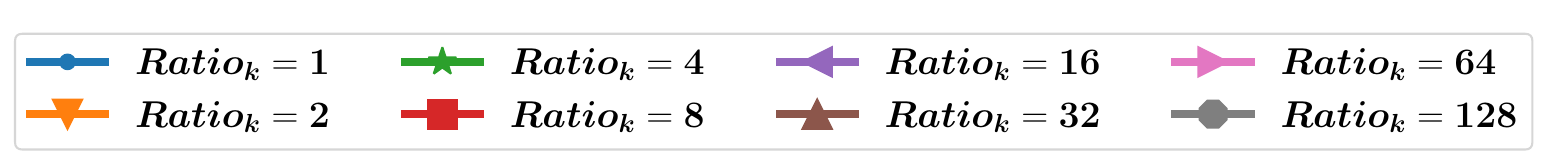}

\includegraphics[width=0.24\linewidth]{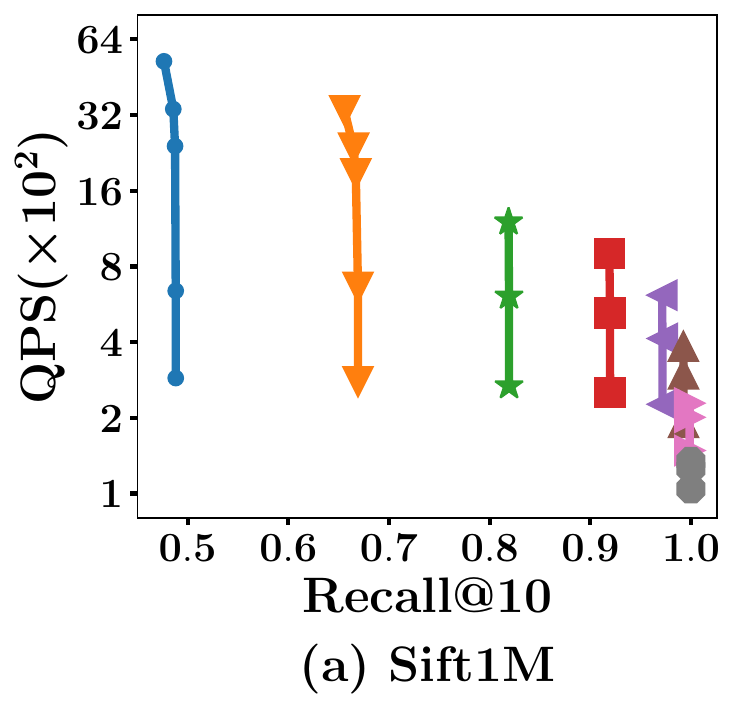}
\includegraphics[width=0.24\linewidth]{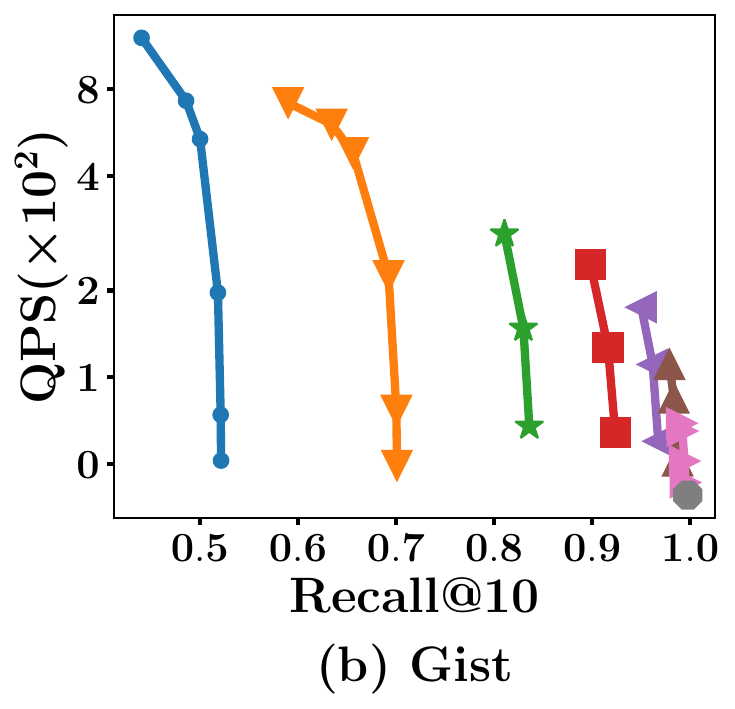}
\includegraphics[width=0.24\linewidth]{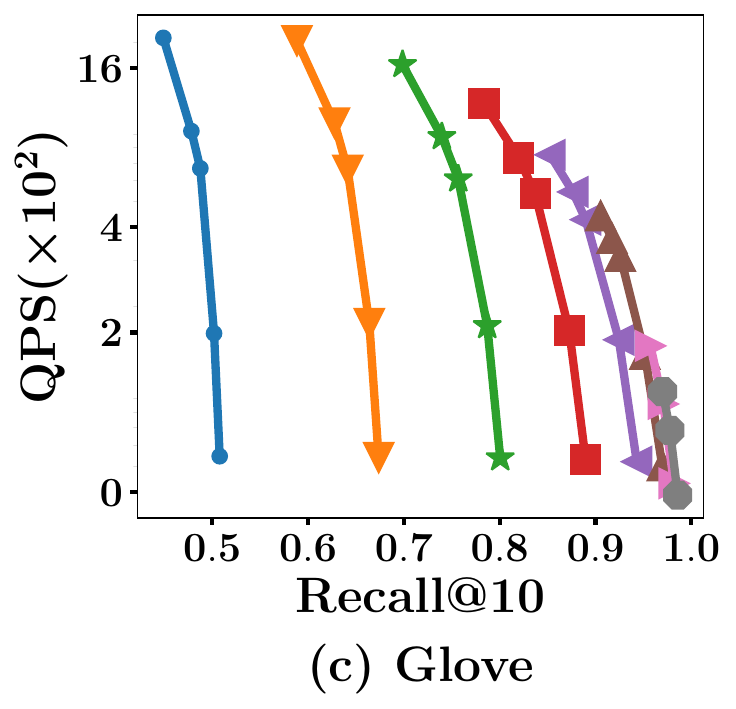}
\includegraphics[width=0.24\linewidth]{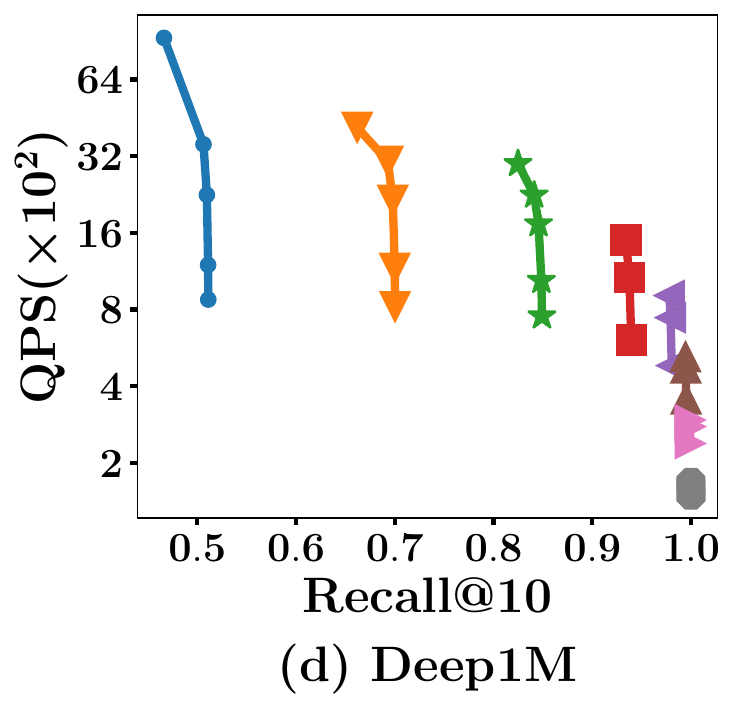}
 \vspace{-4mm}
	\caption{The effects of $Ratio_k$ on the search performance of our solution, where $k^{\prime} = Ratio_k \cdot k$.}
 \vspace{-6mm}
	\label{fig:findk}
\end{figure}

\noindent {\textbf{Compared Methods.}} 
We use the following methods as baselines for comparisons in the experiments. 
\begin{itemize}
    \item \textbf{RS-SANN}~\cite{peng2017reusable}. It encrypts the database via Advanced Encryption Standard (AES) and employs locality-sensitive hashing (LSH) as the index. It requires the user to decrypt selected encrypted candidates from the server and then select the best $k$ neighbors among them.  
    \item \textbf{PACM-ANN}~\cite{zhou2024pacmann}. It uses proximity graphs as the index and employs private information retrieval to obtain graph edges and encrypted vectors from the server in a multi-round manner. The search process is controlled by user. 
    \item \textbf{PRI-ANN}~\cite{servan2022private}. It uses LSH as the index and employs private information retrieval to obtain the candidates from the server. The search process consists of a single round of communication between the client and the two servers. And no communication is required between the servers. \eat{Like PACM-ANN, the user conducts the search process by randomly selecting $k$ candidates from the candidates.}  
\end{itemize}
Besides, we consider a variant of our method denoted \textbf{HNSW-AME} as a baseline. Like our method, HNSW-AME builds an HNSW graph on the encrypted database $C_{DCPE}^P$, but stores AME~\cite{zheng2024achieving} instead of DCE. During the search process, HNSW-AME shares the same \emph{filter} phase but conducts the SDC computations via AME in the \emph{refine} phase. 

\noindent {\textbf{Computing Environment.}} All experiments are conducted on a server equipped with two Intel(R) Xeon(R) Silver 4210R CPUs, each of which has 10 cores, and 256 GB RAM as its main memory. The OS version is CentOS 7.9.2009. All codes are implemented in C++, and the search performance is conducted with a single thread. 

\subsection{Effects of Key Parameters}
\label{ssec:parasel}

In this part, we study the effects of the key parameters of our method on the search performance. 
First, there are key parameters of building HNSW, i.e., $m$ and $efConstruction$. We set $efConstruction=600$ and $m=40$ to build HNSW for each dataset, which is selected by a grid search method. Besides, we vary the search parameter $efSearch$ of HNSW in order to show the balance between efficiency and accuracy. 

\noindent{\textbf{Effects of DCPE Parameters.}}
According to \cite{fuchsbauer2022approximate}, two key parameters in DCPE are $s$ and $\beta$. We set $s=1024$ as recommended in \cite{bogatov2022secure}. $\beta$ should be in the range of $\sqrt{M}$ and $2M\sqrt{d}$, where $M$ indicates the maximum absolute coordinates. Hence, we need to tune $\beta$. 
A larger $\beta$ indicates more noise added and the encrypted vector is further from the plaintext one, which enhances the privacy but degrades the quality of candidates returned in the \emph{filter} phase. Hence, $\beta$ should be carefully tuned. 
We show the influence of $\beta$ on the search performance of our scheme with only \emph{filter} phase in \cref{fig:beta_recall}, where $k^{\prime} = k$. As $\beta$ increases, the upper bound of recall decreases due to more noise added by DCPE. 
In order to balance data privacy and accuracy, we choose the best $\beta$ for each dataset so that the upper bound of recall in \emph{filter} phase is around 0.5. In this case, the attacker's probability of guessing the true neighbor correctly is only $50\%$. Specifically, the best $\beta$ is set as 450, 2.5, 5, and 1.1 for Sift1M, Gist, Glove, and Deep1M,    respectively. 


\noindent{\textbf{Effects of $k^{\prime}$.}}
The number $k^{\prime}$, indicating the number of returned neighbors in \emph{filter} phase, is a key parameter in our scheme. For simplicity, we use the ratio $Ratio_k = k^{\prime} / k$ to imply $k^{\prime}$. As shown in Figure \ref{fig:findk}, we can see that as $Ratio_k$ rises, the upper bound of $Recall@k$ grows but the efficiency decreases. This is because a larger $Ratio_k$ indicates more neighbors verified in \emph{refine} phase, which increases the chance of finding better neighbors but at the expense of more distance comparisons. In the rest of the experiments, we vary $Ratio_k$ for best performance under different $Recall@k$.

\begin{figure}[t]
	\centering
    \includegraphics[width=0.6\linewidth]{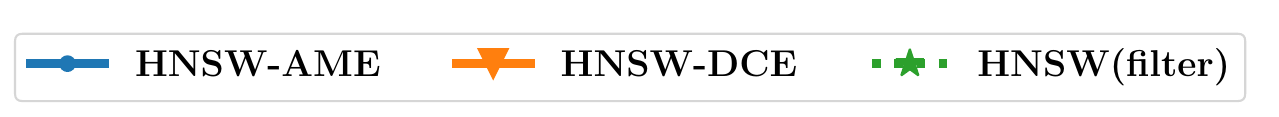}

\includegraphics[width=0.24\linewidth]{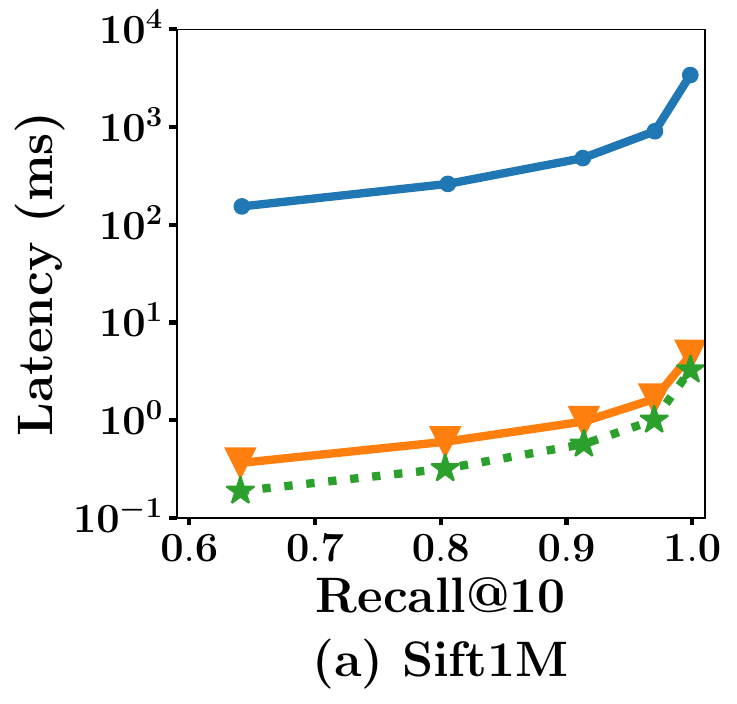}
\includegraphics[width=0.24\linewidth]{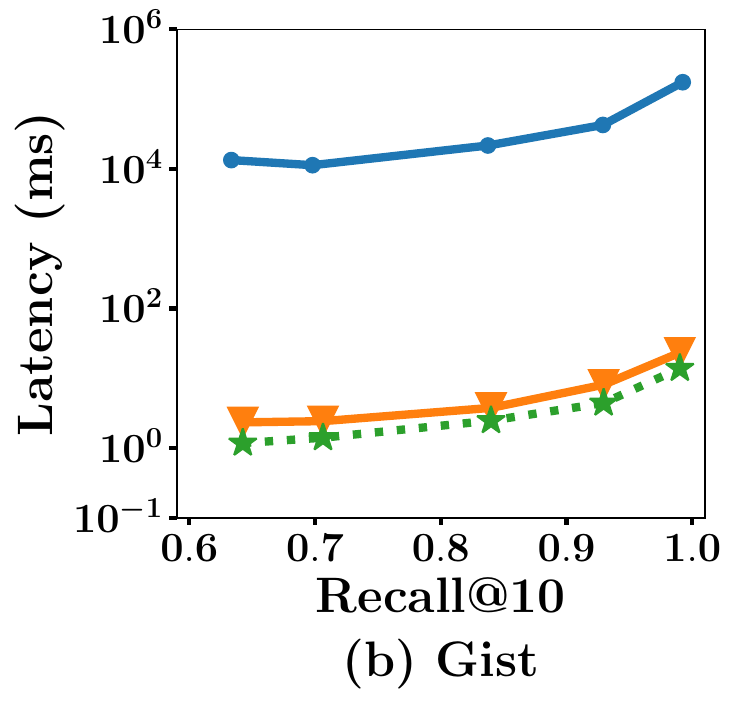}
\includegraphics[width=0.24\linewidth]{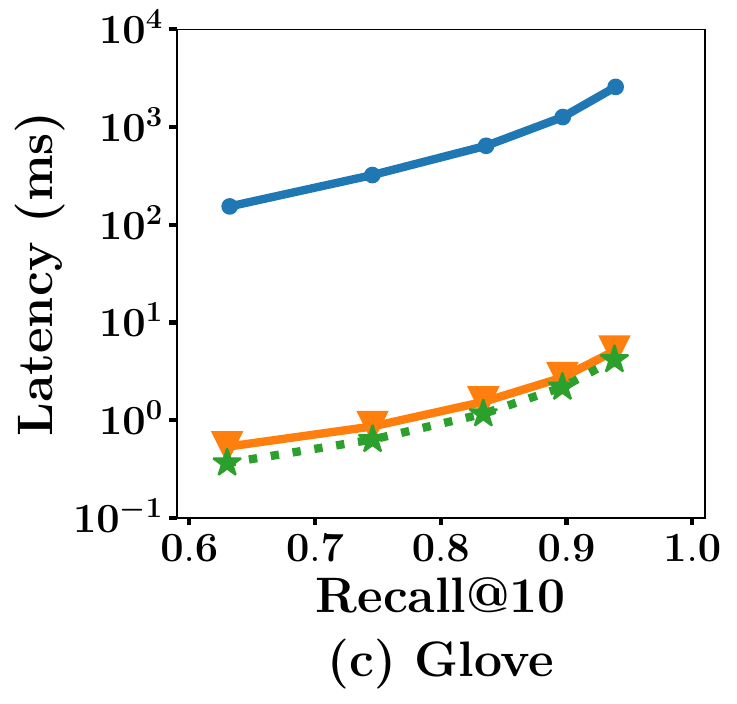}
\includegraphics[width=0.24\linewidth]{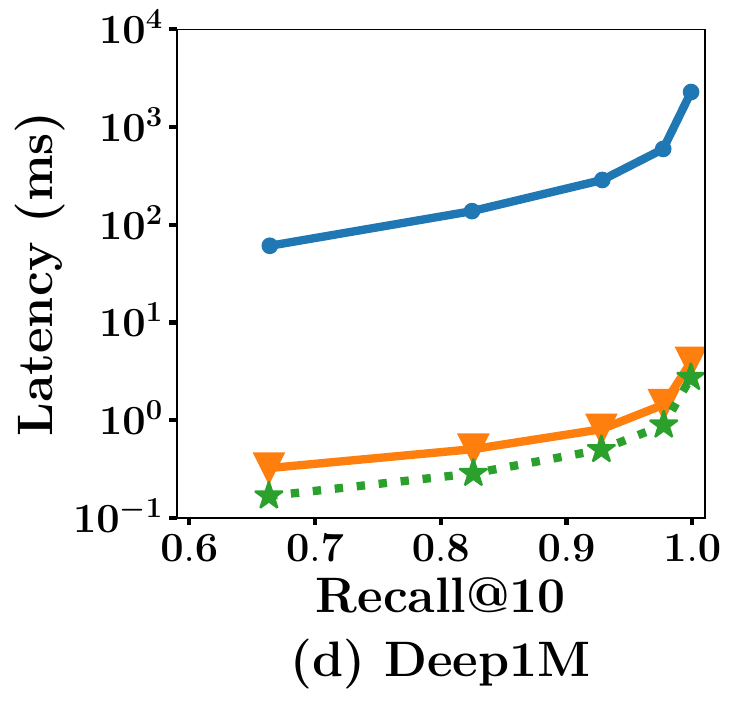}
  \vspace{-6mm}
	\caption{Comparing HNSW-AME with our method denoted as HNSW-DCE. HNSW(filter) indicates our method with only the \emph{filter} phase.}
  \vspace{-3mm}
	\label{fig:ame_dce}
\end{figure}

\begin{figure}[t]
	\centering
    \includegraphics[width=0.8\linewidth]{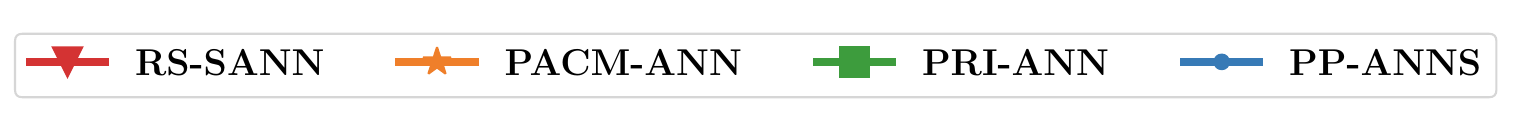}

\includegraphics[width=0.24\linewidth]{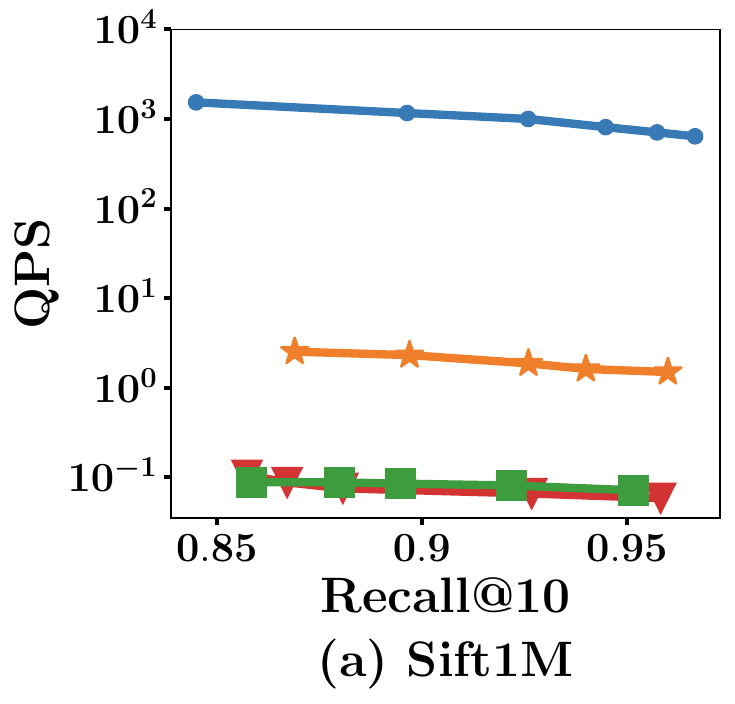}
\includegraphics[width=0.24\linewidth]{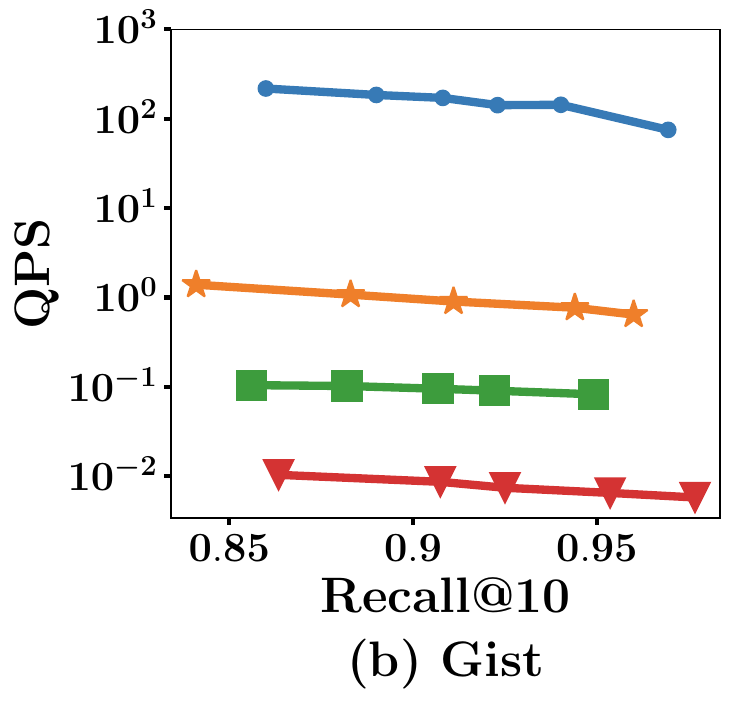}
\includegraphics[width=0.24\linewidth]{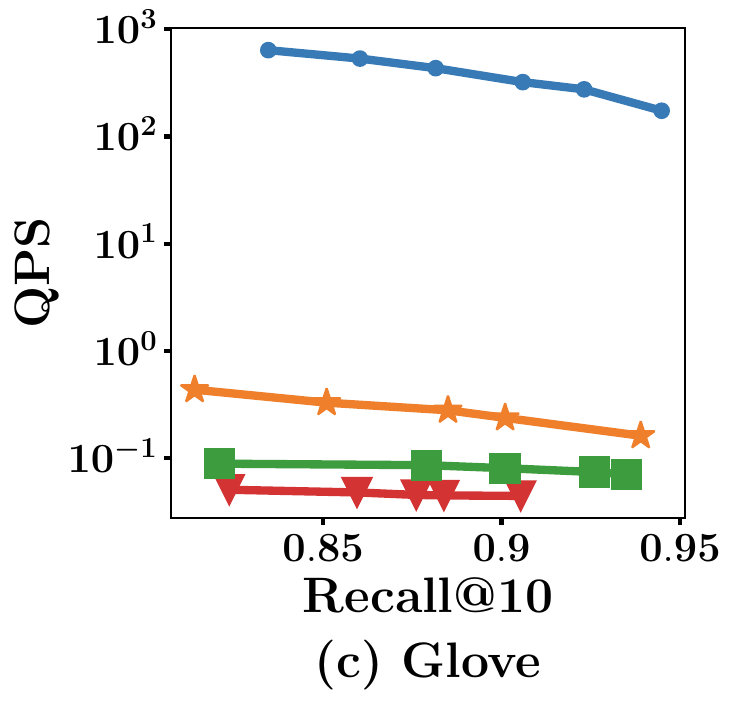}
\includegraphics[width=0.24\linewidth]{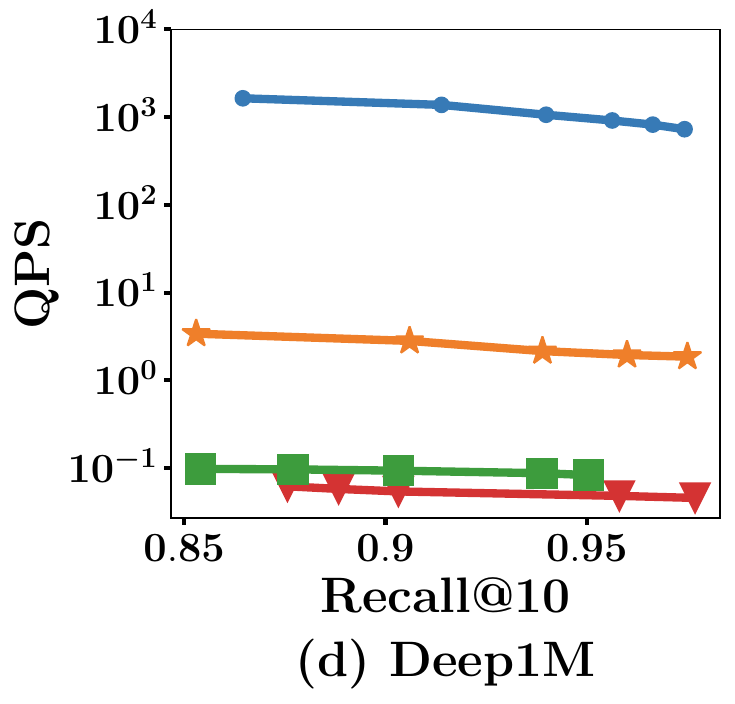}
  \vspace{-6mm}
	\caption{Comparing our method with the baseline methods.}
  \vspace{-3mm}
	\label{fig:cp}
\end{figure}


\begin{figure*}[t]
	\centering
    \begin{minipage}{0.22\linewidth}
	\includegraphics[width=\linewidth]{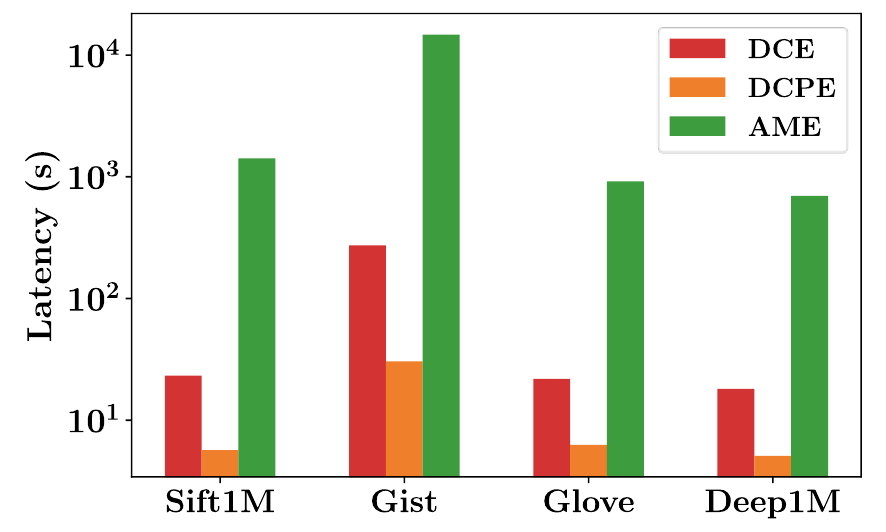}
  \vspace{-4mm}
	\caption{Comparisons in vector encryption cost.}
  \vspace{-3mm}
	\label{fig:cmp_encryption_cost}
        \end{minipage}
        \hspace{1mm}
        \begin{minipage}{0.38\linewidth}
            \includegraphics[width=\linewidth]{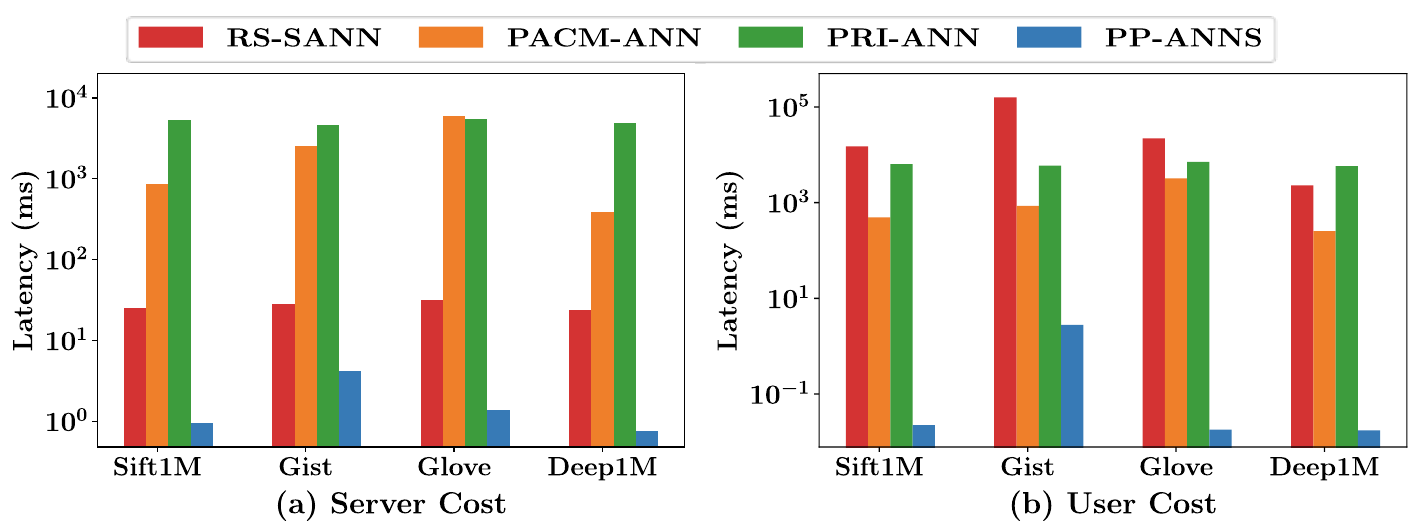}
  \vspace{-6mm}
	\caption{Comparisons in server-side and user-side costs when $Recall@k = 0.9$ (user cost is simulated in the server).}
  \vspace{-3mm}
	\label{fig:comm}
        \end{minipage}
        \hspace{1mm}
        \begin{minipage}{0.35\linewidth}
            \includegraphics[width=\linewidth]{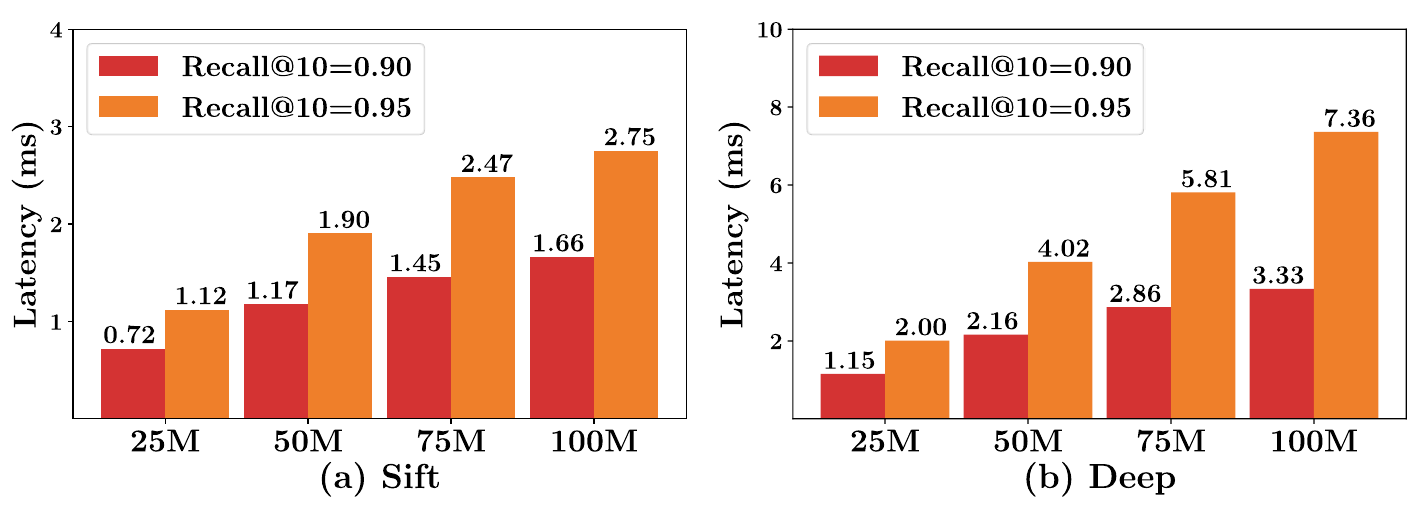}
  \vspace{-6mm}
	\caption{Scalability study of our PP-ANNS scheme.}
  \vspace{-3mm}
	\label{fig:scalability}
        \end{minipage}
\end{figure*}




\subsection{Comparisons with Baseline Methods}
\label{ssec:comp_baseline}

In this part, we compare four baseline methods: HNSW-AME, RS-SANN~\cite{peng2017reusable}, PACM-ANN~\cite{zhou2024pacmann}, and PRI-ANN~\cite{servan2022private}. First, we compare HNSW-AME with our method. 
HNSW-AME has the same parameter settings as ours. Besides, we compare with HNSW(filter) which only has the \emph{filter} phase of ours. 
We show the results in Figure \ref{fig:ame_dce}, where our method is denoted as HNSW-DCE. We can see that HNSW-DCE significantly outperforms HNSW-AME in efficiency, and achieves at least $100\times$ speedup over HNSW-AME. Note that both of them share the same \emph{filter} phase but distinct \emph{refine} phase with different SDC methods. Notably, each SDC of DCE costs $O(d)$, while that of AME costs $O(d^2)$. On the other hand, the latency of HNSW-DCE is pretty close to that of HNSW(filter) for the same accuracy, which indicates that the \emph{refine} phase of our method is very efficient. 

{
Notably, the data provider needs to encrypt the plaintext vectors by DCPE, DCE or AME, and thus consumes significant cost. Here, we compare the encryption cost of the vectors by different encryption methods in Figure~\ref{fig:cmp_encryption_cost}. We can see that AME consumes considerably more cost than DCE, while DCPE costs the least due to its simple encryption method. }



Moreover, we compare our method with the other three baseline methods, as shown in Figure~\ref{fig:cp} and Figure~\ref{fig:comm}. We can see that our method significantly outperforms its competitors in search performance, and has considerably less cost on both the server side and user side.  
RS-SANN follows \emph{filter-and-refine} strategy and conducts \emph{filter} phase in the server, but performs the \emph{refine} phase on the user side, which leads to numerous user-side costs. Besides, it uses LSH as the index and has to retrieve many more candidates to reach the same accuracy as ours. 
Like ours, PACM-ANN employs a proximity graph as the index but conducts the search process on the user side, which interactively fetches index data and database vectors from the server. Hence, it suffers from heavy computational costs on the user side and communication overhead. 
Like RS-SANN, PRI-ANN employs LSH as the index, which leads to numerous candidates for high accuracy. It incurs heavy computational consumption for servers and users. \eat{But it uses two servers for private information retrieval and thus incurs heavy communication overhead between two servers. }
In addition, we conduct experiments for $k$-ANNS with HNSW on plaintext vectors. Compared with it, our PP-ANNS scheme consumes $5\times$, $7\times$, $3\times$, and $4\times$ costs when $Recall@k=0.9$. 

\subsection{Scalability of Our Method}
\label{ssec:scalability}

In this part, we study the scalability of our PP-ANNS scheme. We conduct experiments on four random samples of Sift1B and Deep1B with various sizes, including 25 million, 50 million, 75 million, and 100 million vectors respectively. The experimental results are shown in Figure~\ref{fig:scalability}. We can see that the search performance of our method scales well as the data size increases. To be specific, when achieving the same accuracy, the latency of each query sublinearly grows as the data size increases.

\section{Related Works}\label{sec:rel}

There are two problems related to our work: $k$-ANNS and PP-ANNS on plaintext and encrypted databases. 

\noindent
\textbf{$k$-ANNS Methods in Plaintext Database.} There is a bulk of $k$-ANNS methods in the literature, which could be divided into index-based methods and embedding-based methods. The former employs an index structure, such as locality-sensitive hashing\cite{MPlsh,liu2014sk}, inverted files~\cite{jegou2010product, zheng2023learned}, and proximity graphs~\cite{malkov2018efficient, Nsg, zhao2023towards, Vbase, FastPG, SimJoin} to accelerate the search process. On the other hand, the latter embeds vectors into other embeddings such as production quantization codes~\cite{jegou2010product} and hashing codes~\cite{Wang2017PAMI}, and replaces the expensive distance by a fast and approximate distance. 

\noindent
\textbf{PP-ANNS Schemes.}
A number of PP-ANNS methods have been proposed in the last decades. Those methods could be divided into distance incomparable encryption and distance comparable encryption according to the encryption methods used. The former~\cite{peng2017reusable, chen2020sanns, servan2022private, zhou2024pacmann} has to receive encrypted candidate vectors retrieved from the server and then decrypt them for distance comparisons on the user side. Such methods cause heavy user-side involvement and considerable communication overhead. On the other hand, the latter employ secure distance comparison methods such as asymmetric scalar-product-preserving encryption~\cite{wong2009secure,cao2013privacy, wang2014privacy,miao2023efficient} distance-comparison-preserving encryption~\cite{fuchsbauer2022approximate}, asymmetric matrix encryption~\cite{zheng2024achieving} and homomorphic encryption~\cite{xue2017secure, furon2013fast}.

\section{Conclusion}\label{sec:conclu}

In this paper, we study the privacy-preserving $k$-approximate nearest neighbor search (PP-ANNS) problem and design a new scheme that achieves data privacy, high efficiency, high accuracy, and minimal user-side involvement simultaneously. Our scheme conducts the search process in the server upon receiving an encrypted query vector, in order to minimize the user-side involvement. To protect data privacy, we propose a novel encryption method called distance comparison encryption for secure, exact, and efficient distance comparisons. Moreover, we propose a privacy-preserving index by combining HNSW and DCPE, so as to balance data privacy and search performance. 
Moreover, we theoretically analyze the security of our method and conduct extensive experiments to demonstrate its superiority over existing methods.  

\section*{Acknowledgements}
This work was supported by National Natural Science Foundation of China (No. 62002274, 62272369, 62372352) and Research Grants Council of Hong Kong (No.14205520). 

\bibliographystyle{plain}
\bibliography{sample}

\begin{thebibliography}{10}

\bibitem{baeza1999modern}
Ricardo~A. Baeza{-}Yates and Berthier~A. Ribeiro{-}Neto.
\newblock {\em Modern Information Retrieval}.
\newblock {ACM} Press / Addison-Wesley, 1999.

\bibitem{bailey2004efficient}
Donald~G. Bailey.
\newblock An efficient euclidean distance transform.
\newblock In {\em {IWCIA}}, volume 3322 of {\em Lecture Notes in Computer
  Science}, pages 394--408. Springer, 2004.

\bibitem{barndorff1982exponential}
O~Barndorff-Nielsen, Preben Blaesild, J~Ledet Jensen, and B~J{\o}rgensen.
\newblock Exponential transformation models.
\newblock {\em Proceedings of the Royal Society of London. A. Mathematical and
  Physical Sciences}, 379(1776):41--65, 1982.

\bibitem{bogatov2022secure}
Dmytro Bogatov.
\newblock {\em Secure and efficient query processing in outsourced databases}.
\newblock PhD thesis, Boston University, 2022.

\bibitem{cambareri2015known}
Valerio Cambareri, Mauro Mangia, Fabio Pareschi, Riccardo Rovatti, and Gianluca
  Setti.
\newblock On known-plaintext attacks to a compressed sensing-based encryption:
  {A} quantitative analysis.
\newblock {\em {IEEE} TIFS}, 10(10):2182--2195, 2015.

\bibitem{cao2013privacy}
Ning Cao, Cong Wang, Ming Li, Kui Ren, and Wenjing Lou.
\newblock Privacy-preserving multi-keyword ranked search over encrypted cloud
  data.
\newblock {\em {IEEE} TPDS}, 25(1):222--233, 2014.

\bibitem{chen2020sanns}
Hao Chen, Ilaria Chillotti, Yihe Dong, Oxana Poburinnaya, Ilya~P. Razenshteyn,
  and M.~Sadegh Riazi.
\newblock {SANNS:} scaling up secure approximate k-nearest neighbors search.
\newblock In {\em {USENIX} Security}, pages 2111--2128, 2020.

\bibitem{cullen2012matrices}
Charles~G Cullen.
\newblock {\em Matrices and linear transformations}.
\newblock Courier Corporation, 2012.

\bibitem{Nsg}
Cong Fu, Chao Xiang, Changxu Wang, and Deng Cai.
\newblock Fast approximate nearest neighbor search with the navigating
  spreading-out graph.
\newblock {\em PVLDB}, 12(5):461--474, 2019.

\bibitem{fuchsbauer2022approximate}
Georg Fuchsbauer, Riddhi Ghosal, Nathan Hauke, and Adam O'Neill.
\newblock Approximate distance-comparison-preserving symmetric encryption.
\newblock In {\em {SCN} 2022}, volume 13409 of {\em Lecture Notes in Computer
  Science}, pages 117--144. Springer, 2022.

\bibitem{furon2013fast}
Teddy Furon, Herv{\'{e}} J{\'{e}}gou, Laurent Amsaleg, and Benjamin Mathon.
\newblock Fast and secure similarity search in high dimensional space.
\newblock In {\em {WIFS} 2013}, pages 73--78. {IEEE}, 2013.

\bibitem{guan2021toward}
Yunguo Guan, Rongxing Lu, Yandong Zheng, Jun Shao, and Guiyi Wei.
\newblock Toward oblivious location-based k-nearest neighbor query in smart
  cities.
\newblock {\em {IEEE} Internet Things J.}, 8(18):14219--14231, 2021.

\bibitem{jegou2010product}
Herv{\'{e}} J{\'{e}}gou, Matthijs Douze, and Cordelia Schmid.
\newblock Product quantization for nearest neighbor search.
\newblock {\em {IEEE} TPAMI}, 33(1):117--128, 2011.

\bibitem{jordan2015machine}
Michael~I Jordan and Tom~M Mitchell.
\newblock Machine learning: Trends, perspectives, and prospects.
\newblock {\em Science}, 349(6245):255--260, 2015.

\bibitem{keene1995log}
Oliver~N Keene.
\newblock The log transformation is special.
\newblock {\em Statistics in Medicine}, 14(8):811--819, 1995.

\bibitem{kim2018function}
Sam Kim, Kevin Lewi, Avradip Mandal, Hart Montgomery, Arnab Roy, and David~J.
  Wu.
\newblock Function-hiding inner product encryption is practical.
\newblock In {\em {SCN} 2018}, volume 11035 of {\em Lecture Notes in Computer
  Science}, pages 544--562. Springer, 2018.

\bibitem{RAG}
Patrick Lewis, Ethan Perez, Aleksandra Piktus, Fabio Petroni, Vladimir
  Karpukhin, Naman Goyal, Heinrich K{\"u}ttler, Mike Lewis, Wen-tau Yih, Tim
  Rockt{\"a}schel, et~al.
\newblock Retrieval-augmented generation for knowledge-intensive {NLP} tasks.
\newblock {\em NeurIPS}, 33:9459--9474, 2020.

\bibitem{li2019insecurity}
Rui Li, Alex~X. Liu, Ying Liu, Huanle Xu, and Huaqiang Yuan.
\newblock Insecurity and hardness of nearest neighbor queries over encrypted
  data.
\newblock In {\em {ICDE} 2019}, pages 1614--1617. {IEEE}, 2019.

\bibitem{Dpg}
Wen Li, Ying Zhang, Yifang Sun, Wei Wang, Mingjie Li, Wenjie Zhang, and Xuemin
  Lin.
\newblock Approximate nearest neighbor search on high dimensional data --
  experiments, analyses, and improvement.
\newblock {\em IEEE TKDE}, 32(8):1475--1488, 2019.

\bibitem{lin2017revisiting}
Weipeng Lin, Ke~Wang, Zhilin Zhang, and Hong Chen.
\newblock Revisiting security risks of asymmetric scalar product preserving
  encryption and its variants.
\newblock In {\em ICDCS}, pages 1116--1125. IEEE, 2017.

\bibitem{liu2014sk}
Yingfan Liu, Jiangtao Cui, Zi~Huang, Hui Li, and Heng~Tao Shen.
\newblock Sk-lsh: an efficient index structure for approximate nearest neighbor
  search.
\newblock {\em PVLDB}, 7(9):745--756, 2014.

\bibitem{MPlsh}
Qin Lv, William Josephson, Zhe Wang, Moses Charikar, and Kai Li.
\newblock {Multi-probe LSH}: efficient indexing for high-dimensional similarity
  search.
\newblock In {\em VLDB}, pages 950--961, 2007.

\bibitem{malkov2018efficient}
Yury~A. Malkov and Dmitry~A. Yashunin.
\newblock Efficient and robust approximate nearest neighbor search using
  hierarchical navigable small world graphs.
\newblock {\em {IEEE} TPAMI}, 42(4):824--836, 2020.

\bibitem{miao2023efficient}
Yinbin Miao, Yutao Yang, Xinghua Li, Linfeng Wei, Zhiquan Liu, and Robert~H
  Deng.
\newblock Efficient privacy-preserving spatial data query in cloud computing.
\newblock {\em IEEE TKDE}, 2023.

\bibitem{peng2017reusable}
Yanguo Peng, Jiangtao Cui, Hui Li, and Jianfeng Ma.
\newblock A reusable and single-interactive model for secure approximate
  k-nearest neighbor query in cloud.
\newblock {\em Information Sciences}, 387:146--164, 2017.

\bibitem{taumg}
Yun Peng, Byron Choi, Tsz~Nam Chan, Jianye Yang, and Jianliang Xu.
\newblock Efficient approximate nearest neighbor search in multi-dimensional
  databases.
\newblock {\em Proc. {ACM} Manag. Data}, 1(1):54:1--54:27, 2023.

\bibitem{servan2022private}
Sacha Servan-Schreiber, Simon Langowski, and Srinivas Devadas.
\newblock Private approximate nearest neighbor search with sublinear
  communication.
\newblock In {\em S\&P 2022}, pages 911--929. IEEE, 2022.

\bibitem{singh2013study}
Gurpreet Singh.
\newblock A study of encryption algorithms (rsa, des, 3des and aes) for
  information security.
\newblock {\em International Journal of Computer Applications}, 67(19), 2013.

\bibitem{wang2014privacy}
Bing Wang, Shucheng Yu, Wenjing Lou, and Y.~Thomas Hou.
\newblock Privacy-preserving multi-keyword fuzzy search over encrypted data in
  the cloud.
\newblock In {\em {INFOCOM} 2014}, pages 2112--2120. {IEEE}, 2014.

\bibitem{Wang2017PAMI}
Jingdong Wang, Ting Zhang, Jingkuan Song, Nicu Sebe, and Heng~Tao Shen.
\newblock A survey on learning to hash.
\newblock {\em IEEE TPAMI}, 40(4):769--790, 2017.

\bibitem{survey2021}
Mengzhao Wang, Xiaoliang Xu, Qiang Yue, and Yuxiang Wang.
\newblock A comprehensive survey and experimental comparison of graph-based
  approximate nearest neighbor search.
\newblock {\em PVLDB}, 14(11):1964–1978, 2021.

\bibitem{wong2009secure}
Wai~Kit Wong, David Wai-lok Cheung, Ben Kao, and Nikos Mamoulis.
\newblock Secure knn computation on encrypted databases.
\newblock In {\em SIGMOD}, pages 139--152, 2009.

\bibitem{SimJoin}
Jiadong Xie, Jeffrey~Xu Yu, and Yingfan Liu.
\newblock Fast approximate similarity join in vector databases.
\newblock In {\em SIGMOD}. ACM, 2025.

\bibitem{xue2017secure}
Wenzhuo Xue, Hui Li, Yanguo Peng, Jiangtao Cui, and Yu~Shi.
\newblock Secure k nearest neighbors query for high-dimensional vectors in
  outsourced environments.
\newblock {\em IEEE Transactions on Big Data}, 4(4):586--599, 2018.

\bibitem{FastPG}
Shuo Yang, Jiadong Xie, Yingfan Liu, Jeffrey~Xu Yu, Xiyue Gao, Qianru Wang,
  Yanguo Peng, and Jiangtao Cui.
\newblock Revisiting the index construction of proximity graph-based
  approximate nearest neighbor search.
\newblock {\em PVLDB}, 2025.

\bibitem{yao2013secure}
Bin Yao, Feifei Li, and Xiaokui Xiao.
\newblock Secure nearest neighbor revisited.
\newblock In {\em ICDE}, pages 733--744. IEEE, 2013.

\bibitem{yuan2017practical}
Jiawei Yuan and Yifan Tian.
\newblock Practical privacy-preserving mapreduce based k-means clustering over
  large-scale dataset.
\newblock {\em {IEEE} Transactions on Cloud Computing}, 7(2):568--579, 2019.

\bibitem{Vbase}
Qianxi Zhang, Shuotao Xu, Qi~Chen, Guoxin Sui, Jiadong Xie, Zhizhen Cai, Yaoqi
  Chen, Yinxuan He, Yuqing Yang, Fan Yang, Mao Yang, and Lidong Zhou.
\newblock {VBASE:} unifying online vector similarity search and relational
  queries via relaxed monotonicity.
\newblock In {\em OSDI}, pages 377--395. {USENIX} Association, 2023.

\bibitem{zhang2018secure}
Zhilin Zhang, Ke~Wang, Chen Lin, and Weipeng Lin.
\newblock Secure top-k inner product retrieval.
\newblock In {\em {CIKM} 2018}, pages 77--86. {ACM}, 2018.

\bibitem{zhao2023towards}
Xi~Zhao, Yao Tian, Kai Huang, Bolong Zheng, and Xiaofang Zhou.
\newblock Towards efficient index construction and approximate nearest neighbor
  search in high-dimensional spaces.
\newblock {\em PVLDB}, 16(8):1979--1991, 2023.

\bibitem{zheng2023learned}
Bolong Zheng, Ziyang Yue, Qi~Hu, Xiaomeng Yi, Xiaofan Luan, Charles Xie,
  Xiaofang Zhou, and Christian~S Jensen.
\newblock Learned probing cardinality estimation for high-dimensional
  approximate nn search.
\newblock In {\em ICDE}, pages 3209--3221. IEEE, 2023.

\bibitem{zheng2018efficient}
Yandong Zheng and Rongxing Lu.
\newblock An efficient and privacy-preserving $k$-nn query scheme for
  ehealthcare data.
\newblock In {\em Physical and Social Computing (CPSCom) and {IEEE} Smart Data
  (SmartData)}, pages 358--365. {IEEE}, 2018.

\bibitem{zheng2019achieving}
Yandong Zheng, Rongxing Lu, and Jun Shao.
\newblock Achieving efficient and privacy-preserving k-nn query for outsourced
  ehealthcare data.
\newblock {\em Journal of Medical Systems}, 43:1--13, 2019.

\bibitem{zheng2024achieving}
Yandong Zheng, Rongxing Lu, Songnian Zhang, Jun Shao, and Hui Zhu.
\newblock Achieving practical and privacy-preserving knn query over encrypted
  data.
\newblock {\em IEEE TDSC}, 2024.

\bibitem{zhou2024pacmann}
Mingxun Zhou, Elaine Shi, and Giulia Fanti.
\newblock Pacmann: Efficient private approximate nearest neighbor search.
\newblock {\em {IACR} Cryptol. ePrint Arch.}, page 1600, 2024.

\bibitem{zhu2013secure}
Youwen Zhu, Rui Xu, and Tsuyoshi Takagi.
\newblock Secure k-nn computation on encrypted cloud data without sharing key
  with query users.
\newblock In {\em SCC@ASIACCS}, pages 55--60. {ACM}, 2013.

\end{thebibliography}

\end{document}